\documentclass[namedreferences]{SolarPhysics}
\usepackage[optionalrh]{spr-sola-addons} 
\usepackage{enumerate}
\usepackage{graphicx}
\usepackage{color}
\usepackage{url}             

\ifx \
url    \undefined \def
\doiurl#1{\href{http://dx.doi.org/#1}{\textsf{DOI}}}\fi \ifx
\adsurl \undefined \def
\adsurl#1{\href{http://adsabs.harvard.edu/abs/#1}{\textsf{ADS}}}\fi
\ifx \arxivurl  \undefined \def
\arxivurl#1{\href{http://arxiv.org/abs/#1}{\textsf{arXiv}}}\fi

\begin{document}
\begin{article}
\begin{opening}

\title{Radio, Hard X-Ray, and Gamma-Ray Emissions Associated with
a Far-Side Solar Event}

\author{\inits{V.V.}\fnm{V.}~\lnm{Grechnev}$^{1}$\orcid{0000-0001-5308-6336}}\sep
 \author{\inits{V.I.}\fnm{V.}~\lnm{Kiselev}$^{1}$}\sep
  \author{\inits{L.K.}\fnm{K.}~\lnm{Kashapova}$^{1, 2}$\orcid{0000-0003-2074-5213}}\sep
  \author{\inits{A.A.}\fnm{A.}~\lnm{Kochanov}$^{1, 2}$}\sep
  \author{\inits{I.V.}\fnm{I.}~\lnm{Zimovets}$^{3, 4, 5}$}\sep
  \author{\inits{A.M.}\fnm{A.}~\lnm{Uralov}$^{1}$}\sep
  \author{\inits{B.A.}\fnm{B.}~\lnm{Nizamov}$^{6, 7}$}\sep
  \author{\inits{I.Yu.}\fnm{I.}~\lnm{Grigorieva}$^{8}$}\sep
  \author{\inits{D.V.}\fnm{D.}~\lnm{Golovin}$^{3}$}\sep
  \author{\inits{M.L.}\fnm{M.}~\lnm{Litvak}$^{3}$}\sep
  \author{\inits{I.G.}\fnm{I.}~\lnm{Mitrofanov}$^{3}$}\sep
  \author{\inits{A.B.}\fnm{A.}~\lnm{Sanin}$^{3}$}

\institute{$^{1}$ Institute of Solar-Terrestrial Physics SB RAS,
             Lermontov St.\ 126A, Irkutsk 664033, Russia
             email: \url{grechnev@iszf.irk.ru} \\
             $^{2}$ Irkutsk State University, Gagarin Blvd.\ 20,
             Irkutsk 664003, Russia \\
             $^{3}$ Space Research Institute RAS,
             Profsoyuznaya St.\ 84/32, Moscow 117997, Russia
             email: \url{ivanzim@iki.rssi.ru} \\
             $^{4}$ State Key Laboratory of Space Weather, National
             Space Science Center (NSSC) of the Chinese Academy of
             Sciences, No.~1 Nanertiao, Zhongguancun, Haidian District, Beijing,
             100190, China \\
             $^{5}$ International Space Science Institute -- Beijing
             (ISSI-BJ), No.~1
             Nanertiao, Zhongguancun, Haidian District, Beijing, 100190,
             China \\
             $^{6}$ Faculty of Physics, Lomonosov Moscow State
             University, Moscow, 119992 Russia \\
             $^{7}$ Sternberg Astronomical Institute, Lomonosov
             Moscow State University, Universitetskii Pr. 13, Moscow,
             119992 Russia \\
             $^{8}$ Main Astronomical (Pulkovo) Observatory RAS,
             Pulkovskoe Sh.\ 65, St. Petersburg 196140, Russia
             email: \url{irina.2014.irina@mail.ru}}

 \runningauthor{V.V. Grechnev \textit{et al.}}
 \runningtitle{Emissions from a Far-Side Solar Event}

\date{Received ; accepted }

\begin{abstract} The SOL2014-09-01 far-side solar
eruptive event produced hard electromagnetic and radio emissions
observed with detectors at near-Earth vantage points. Especially
challenging was a long-duration $>100$\,MeV $\gamma$-ray burst
probably produced by accelerated protons exceeding 300\,MeV. This
observation raised a question of how high-energy protons could reach
the Earth-facing solar surface. Some preceding studies discussed a
scenario in which protons accelerated by a CME-driven shock high in
the corona return to the solar surface. We continue with the
analysis of this challenging event, involving radio images from the
\textit{Nan{\c c}ay Radioheliograph} and hard X-ray data from the
\textit{High Energy Neutron Detector} (HEND) of the
\textit{Gamma-Ray Spectrometer} onboard the \textit{Mars Odyssey}
space observatory located near Mars. HEND recorded unocculted flare
emission. The results indicate that the emissions observed from the
Earth's direction were generated by flare-accelerated electrons and
protons trapped in static long coronal loops. Their reacceleration
is possible in these loops by a shock wave, which was excited by the
eruption, being initially not CME-driven. The results highlight the
ways to address remaining questions.
\end{abstract}

\keywords{Flares, Energetic Particles; Magnetic Fields, Corona;
Radio Bursts; Waves, Shock; X-Ray Bursts}

\end{opening}


\section{Introduction}
  \label{S-introduction}

The source of solar energetic particles (SEPs) produced in solar
eruptive-flare events is a subject of long-standing debate. SEPs
consist of different species dominated by protons. Two sources of
accelerated protons have been considered (see, \textit{e.g.},
\citealp{Kahler2001, Kallenrode2003, Aschwanden2012, Reames2013}).
One presumable origin of accelerated protons is associated with
flare processes in solar active regions manifested in X-rays and
microwaves. Another source is related to a bow-shock driven by the
outer surface of a super-Alfv{\'e}nic coronal mass ejection (CME).
Many indications have been considered to identify the elusive source
of accelerated protons. One of them is $\gamma$-ray emission, which
was mostly observed concurrently with other flare emissions and
seemingly favored the acceleration of protons in flares along with
electrons (\textit{e.g.} \citealp{RamatyMandzhavidze2000,
LivshitsBelov2004, ChuppRyan2009, Kurt2010, Vilmer2011}).

Flare emissions are observed in a wide electromagnetic range, from
radio waves up to high-energy $\gamma$-rays. Gyrosynchrotron
emission observed in the radio range and a broadband hard X-ray
(HXR) and $\gamma$-ray bremsstrahlung continuum are produced by
accelerated electrons. Accelerated protons and heavier ions can be
recognized from discrete $\gamma$-ray lines. Of special interest is
the $\pi^0$-decay emission. Neutral pions appear in proton--proton
collisions, when the proton energy exceeds 300\,MeV, and they
rapidly decay into two photons, producing a Doppler-broadened wide
enhancement in the $\gamma$-ray spectrum around 100\,MeV on top of
the bremsstrahlung continuum (\textit{e.g.}
\citealp{RamatyKozlovskyLingenfelter1975, HudsonRyan1995,
Vilmer2011}). Thus, the $\pi^0$-decay emission is a direct
indication of protons accelerated to high energies. Observations and
identification of this $\gamma$-ray emission are only possible with
a high sensitivity and sufficient spectral measurements at high
energies. For this reason, the total number of events with a
confident identification of the $\pi^0$-decay emission was fewer
than 20 in the past (\textit{e.g.} \citealp{Ryan2000, Grechnev2008,
ChuppRyan2009, Kurt2010, Kuznetsov2011}).

Being temporally close to flare emissions produced by accelerated
electrons, discrete nuclear $\gamma$-ray lines, and especially the
$\pi^0$-decay emission, have been considered as evidence of proton
acceleration in flares. On the other hand, $\gamma$-ray emission
much longer than the HXR burst was observed in a few events
(\textit{e.g.} \citealp{Forrest1985, Akimov1996, Ryan2000}). A
challenge to the flare-related origin of $\gamma$-ray emission was
provided by the observation of $\gamma$-ray emission from an event
behind the solar limb. To explain this phenomenon, \cite{Cliver1993}
proposed that protons accelerated by a CME-driven shock wave on an
open magnetic field partly escaped into interplanetary space and
partly returned to the solar surface, precipitating far from the
flare region.

With the advent of the \textit{Fermi Gamma-Ray Space Telescope}
mission in 2008, high-sensitivity $\gamma$-ray observations become
available with a comprehensive spectral information and coordinate
measurements of $\gamma$-ray photons at $> 100$\,MeV by the
\textit{Large Area Telescope} (LAT: \citealp{Atwood2009}). Although
it is a non-solar mission, \textit{Fermi} also provides rich
information for solar studies. \textit{Fermi} has shown that
$\gamma$-ray emissions are quite common in solar flares. Thirty
long-duration $\gamma$-ray events have been observed
\citep{Share2017}. \cite{Pesce-Rollins2015} reported on the
detection by \textit{Fermi}/LAT of high-energy $\gamma$-ray
emissions from three behind-the-limb solar flares on 11 October
2013, 6 January 2014, and 1 September 2014. These events were
addressed by \cite{Ackermann2017}. The $\pi^0$-decay emission was
identified with confidence in two of them, SOL2013-10-11 and
SOL2014-09-01. The authors invoked the idea of \cite{Cliver1993} to
interpret these emissions.

\cite{Plotnikov2017} elaborated on this idea in their analysis of
the three events. Among the issues analyzed, by means of
three-dimensional reconstructions of coronal shock fronts, the
authors showed the events' magnetic connectivity to the Earth-facing
solar surface. \cite{Ackermann2017} and \cite{Plotnikov2017}
considered coronal shocks to be driven by fast CMEs and emphasized
that the CME and associated shock wave were fastest in the 1
September 2014 event, where the high-energy $\gamma$-ray emission
was strongest.

On the other hand, \cite{Hudson2017} pointed out basic problems of
the scenario proposed by \cite{Cliver1993}. First, a large mirror
ratio at the base of an open coronal structure prevents the
back-precipitation of particles from large coronal heights, so that
only a small part of the protons is able to return to the Sun in
this scheme. Second, the total number of high-energy protons
estimated for a set of SEP events appears to be insufficient to
sustain the high-energy $\gamma$-rays in the events addressed by
\cite{Ackermann2017}.

To explain the long-duration $\gamma$-rays from occulted events,
\cite{Hudson2017} considered two options. In the ``Lasso'' scenario,
some SEPs are captured in a noose, which extends to several solar
radii and then retracts. In this scenario, trapped particles acquire
energy due to the betatron acceleration and first-order Fermi
process. The second option that he proposed is a ``coronal thick
target'' scenario, in which protons trapped in a static volume
generate pions and $\gamma$-ray continuum. As \cite{Hudson2017}
estimated, this can proceed for a few hours.

Analyzing the dynamic evolution of the global magnetic field and the
shock wave considered to be CME-driven, \cite{Jin2018} simulated the
CME in the 1 September 2014 event by using a global MHD model. The
authors concluded that particles responsible for the high-energy
$\gamma$-ray emission were accelerated in the CME environment and
escaped the shock downstream region along magnetic fields connected
to the solar surface far away from the flaring region.

So one has to conclude that, in spite of the rich observational data
available at present and a lot of efforts applied, the source of
accelerated protons escapes identification. Furthermore, examining
the ``flare \textit{vs.} CME-driven shock'' alternative, the
researchers base their considerations on a simplified traditional
scheme of the bow-shock excitation by the outer surface of a fast
CME. However, recent studies show that coronal shock waves are
initially excited by sharply erupting flux-ropes inside the
developing CMEs, while reconnection processes underneath produce a
flare (see, \textit{e.g.}, \citealp{Grechnev2016, Grechnev2018} for
details and review). The flare, CME, and shock-formation processes
turn out to be tightly associated, which determines a close relation
between the characteristics of flares, CMEs, and shock waves. The
situation gets still more complicated.

If flare processes are actually responsible for acceleration of
protons, then parameters of CMEs are misleading. If
shock-acceleration is at work, then the acceleration starts
earlier and at lower altitudes than assumed previously. If both
sources are implicated, then untangling their contributions is
still more difficult.

Keeping in mind these circumstances, we continue with the analysis
of the 1 September 2014 event. It was observed from different
vantage points. From the Earth's direction, the HXR burst was
observed by the \textit{Fermi Gamma-ray Burst Monitor} (GBM:
\citealp{Meegan2009}) and the \textit{Wind}/Konus \textit{Gamma-Ray
Burst Experiment} \citep{Aptekar1995}. A radio burst dominated by
the gyrosynchrotron (GS) emission at frequencies $> 300$\,MHz was
recorded by the \textit{Radio Solar Telescope Network} (RSTN:
\citealp{Guidice1979, Guidice1981}), while its source was observed
at the \textit{Nan{\c c}ay Radioheliograph} (NRH:
\citealp{Kerdraon1997}). The GS burst was considered by
\cite{Ackermann2017} and \cite{Carley2017}. The unocculted flare
emission was recorded from the Martian direction by the \textit{High
Energy Neutron Detector} (HEND) of the \textit{Gamma-Ray
Spectrometer} onboard the \textit{Mars Odyssey} space observatory
\citep{Boynton2004}. The SOL2014-09-01 event was not listed in the
HEND catalog \citep{Livshits2017}; nevertheless, HEND actually
observed it.

Based on these data, we analyze the electromagnetic emissions
observed, endeavor to figure out their possible sources, try to
understand the causes of the long-lasting emissions, and reveal the
history and possible role of the shock wave. In this way, we pursue
understanding which scenarios of those proposed match the
observations, specifying and refining some of the results and
conclusions drawn previously.

Section~\ref{S-emissions_sources} addresses electromagnetic
emissions observed in the event and their probable sources.
Section~\ref{S-shock_wave} analyzes shock waves and their
kinematics. Section~\ref{S-discussion} discusses the results and
their interpretation. Section~\ref{S-conclusion} summarizes the
findings and presents the conclusion.

\section{Electromagnetic Emissions and Their Sources}
 \label{S-emissions_sources}

\subsection{Overview of the Event}
 \label{S-overview}

The eruptive flare occurred in an active region (AR) located behind
the east limb at a position of N14\,E126 estimated by
\cite{Ackermann2017} or N14\,E129 according to our estimate. The AR
was numbered 12158 when it became visible from Earth. The flare was
visible from different vantage points. It was observed from the
STEREO-B spacecraft of the twin \textit{Solar Terrestrial Relations
Observatory} (STEREO: \opencite{Kaiser2008}). STEREO-B was located
$161^\circ$ eastward from Earth. The vantage point of HEND onboard
the \textit{Mars Odyssey} was located $65.3^\circ$ eastward from
Earth.

\begin{figure} 
  \centerline{\includegraphics[width=\textwidth]
   {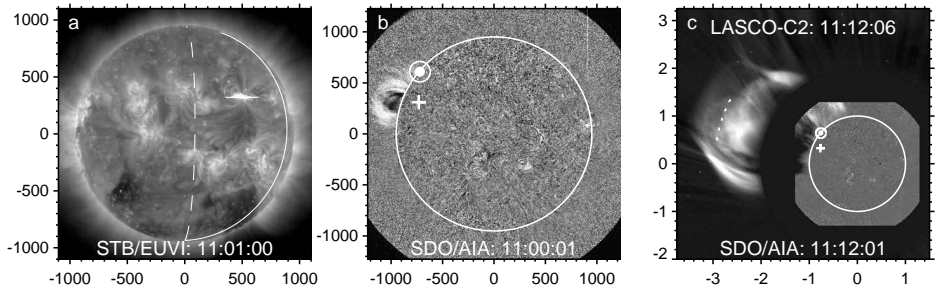}
  }
\caption{The SOL2014-09-01 event observed from different vantage
points. a)~The flare observed by STEREO-B/EUVI in 195\,\AA. The
meridian outlined with solid arc is at a heliolongitude of
$-90^{\circ}$ approximately corresponding to the east limb visible
from Earth. The dashed arc at $-155.3^{\circ}$ corresponds to the
east limb visible from Mars. b)~The early CME lift-off in an SDO/AIA
211\,\AA\ image ratio. The cross denotes the projected position of
the flare site. c)~CME in a LASCO-C2 image with an inserted
co-temporal AIA 211\,\AA\ image ratio. The dotted arc outlines the
CME core at position angles $73^\circ \pm 10^\circ$ according to the
measurements presented in Figure~\ref{F-blob}f. The dot in panels b
and c denotes the \textit{Fermi}/LAT $> 100$\,MeV emission centroid
position with the 68\,\% error circle (from
\citealp{Ackermann2017}). The axes indicate the distance from solar
disk center in arcseconds in panels a and b and in solar radii in
panel c.}
  \label{F-overview}
  \end{figure}

Figure~\ref{F-overview}a presents the flare (bright streak) as
observed by 195\,\AA\ by the \textit{Extreme Ultraviolet Imager}
(EUVI: \opencite{Howard2008}) onboard STEREO-B. As the figure shows,
the flare emission was unocculted for STEREO-B and HEND. According
to \cite{Plotnikov2017}, the flare started in soft X-rays at about
10:54 and peaked at about 11:11 (all times henceforth are adjusted
to observations from 1\,AU and referred to UTC). The GOES importance
of the flare estimated indirectly from STEREO-B/EUVI 195\,\AA\ data
ranged from X1.0 \citep{Chertok2015} to X2.4 estimated by
\cite{Ackermann2017} using the method of \cite{Nitta2013}.

Figure~\ref{F-overview}b exemplifies the observations by the
\textit{Atmospheric Imaging Assembly} (AIA: \citealp{Lemen2012})
onboard the \textit{Solar Dynamic Observatory} (SDO) located at a
near-Earth vantage point. For the analysis we mostly use the
quarter-resolution level 1.5 synoptic AIA data available at
\url{jsoc.stanford.edu/data/aia/synoptic/} in steps of two minutes.
The AIA 211\,\AA\ image ratio in Figure~\ref{F-overview}b presents
the early lift-off of the CME. The projected position of the
far-side active region is denoted by the cross. The dot denotes the
\textit{Fermi}/LAT $> 100$\,MeV emission centroid position with the
68\,\% error circle measured by \cite{Ackermann2017}. The error
circle characterizes the measurement accuracy and should not be
confused with the scatter in the positions of individual
$\gamma$-ray photons, which occupy a very large area of several
solar disks. The \textit{Fermi}/LAT centroid position is commented
on in Section~\ref{S-gamma-ray_source}.

\begin{figure} 
  \centerline{\includegraphics[width=0.8\textwidth]
   {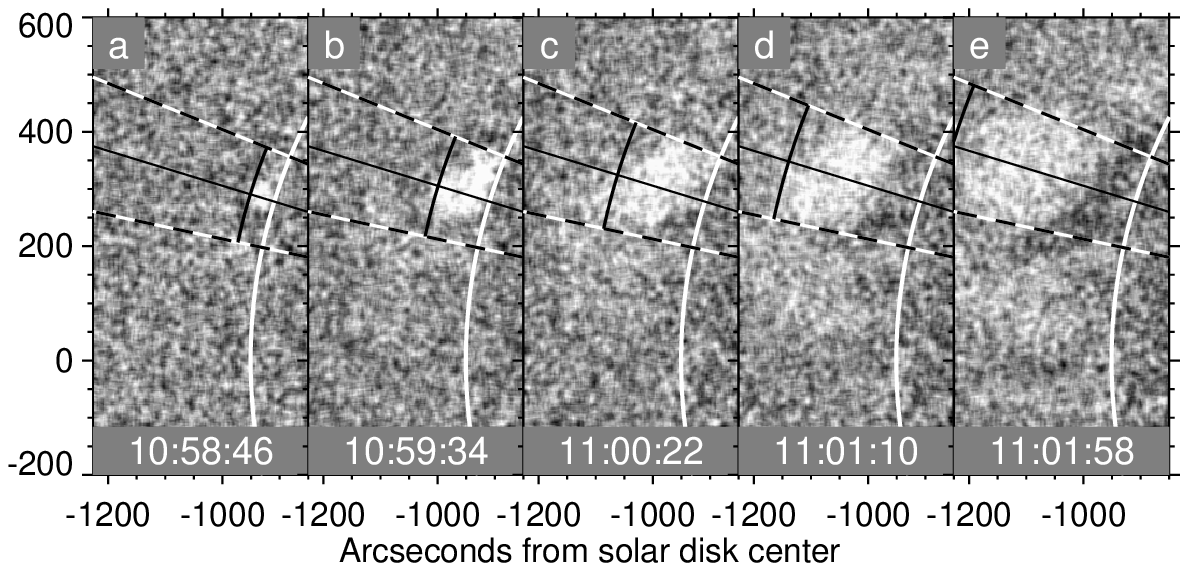}
  }
\vspace{0.1cm}
  \centerline{\includegraphics[width=0.7\textwidth]
   {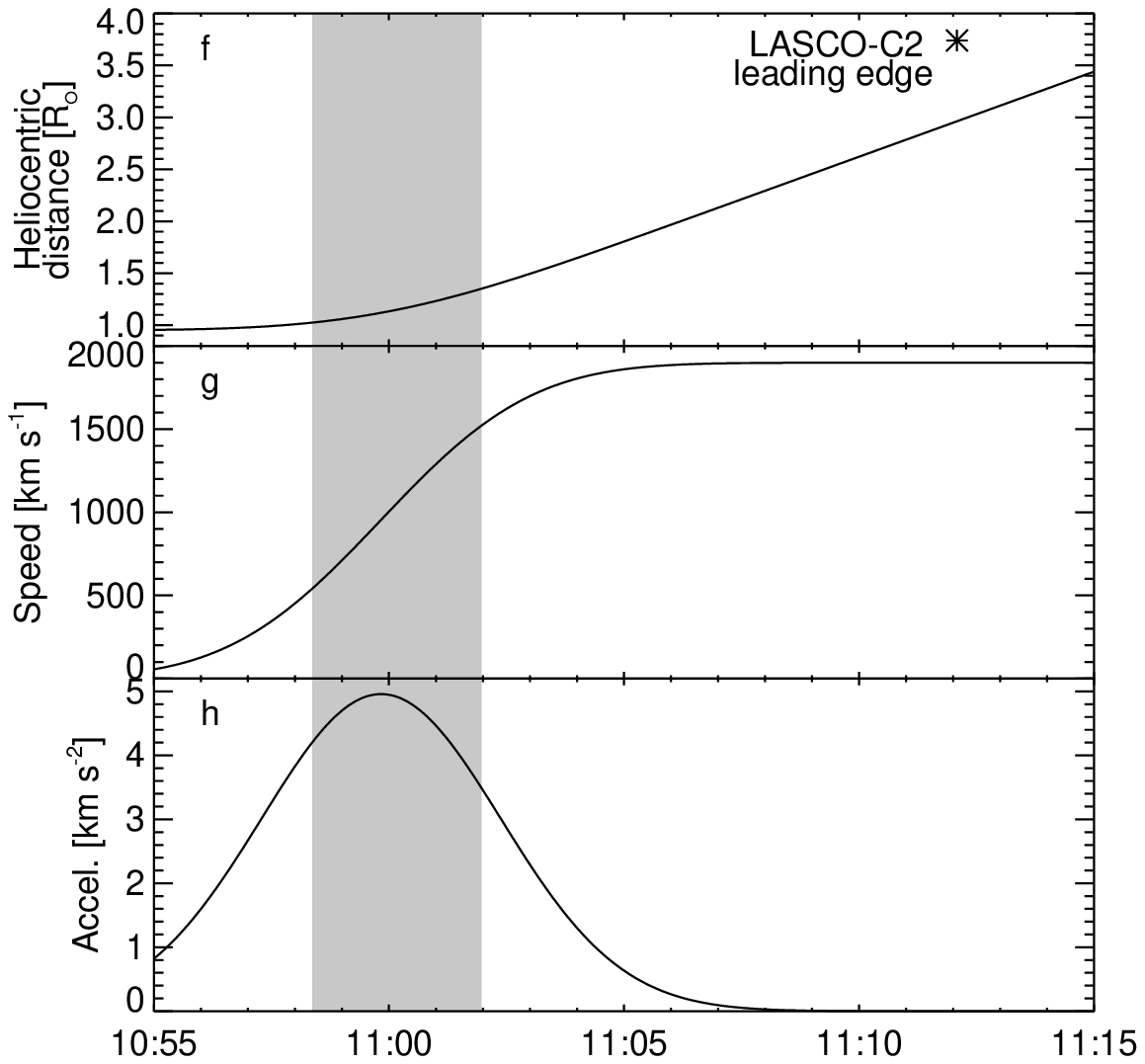}
  }
  \caption{a\,--\,e)~Rising blob in AIA 131\,\AA\ image ratios
(each divided by a fixed pre-event image observed at 10:58:22). The
white circles denote the solar limb. The black arcs outline the
leading edge of the blob. The dashed lines bound the angular extent
of the blob $73^\circ \pm 5^\circ$ with a central position angle
denoted by the straight black line. f\,--\,h)~Probable kinematical
plots of the blob. The asterisk in panel f represents the first CME
measurement in the CME catalog. The shading represents the interval
where the blob was measured within the AIA field of view.}
  \label{F-blob}
  \end{figure}

Figure~\ref{F-overview}c shows the CME observed by the \textit{Large
Angle Spectroscopic Coronagraph} (LASCO: \citealp{Brueckner1995})
onboard the \textit{Solar and Heliospheric Observatory} (SOHO) with
an inserted co-temporal 211\,\AA\ image ratio. The dotted arc
outlines the CME core at position angles $73^\circ \pm 10^\circ$
according to the measurements presented in Figure~\ref{F-blob}f. The
average speed of a fastest CME feature measured in the online CME
catalog (\url{cdaw.gsfc.nasa.gov/CME_list/}: \citealp{Yashiro2004})
at position angles from $76^\circ$ to $60^\circ$ was about
1900\,km\,s$^{-1}$ with a strong average deceleration of
$-240$\,m\,s$^{-2}$. These properties indicate that the measurements
in the CME catalog are related to a shock wave
\citep{Grechnev2011_I}.

It is difficult to detect any erupting feature in EUVI 195\,\AA\
images, whereas rare imaging in different EUVI channels missed the
event. Nevertheless, the AIA 131\,\AA\ images in Figures
\ref{F-blob}a\,--\,\ref{F-blob}e reveal a blob rising radially from
behind the limb. The dashed lines bound the angular extend of the
blob $73^\circ \pm 5^\circ$ with a central position angle denoted by
the straight black line. After an apparent fast initial
three-dimensional expansion, the blob did not exceed laterally the
dashed lines by 11:02:00.

The blob is only visible in 131\,\AA\ and not detectable in any
other channels. The temperature sensitivity characteristics of the
AIA channels \citep{Lemen2012} thus suggest a blob temperature of
about 10\,MK. Most likely, this was an erupting flux rope. Hot flux
ropes have previously been observed in 131\,\AA\ (\textit{e.g.}
\citealp{Cheng2011, Zimovets2012, PatsourakosVourlidasStenborg2013,
Grechnev2016}). The structure of the blob is indiscernible;
nevertheless, the AIA observations allow us to infer its kinematics.

Figures \ref{F-blob}f\,--\,\ref{F-blob}h present probable
kinematical plots of the blob inferred from the AIA 131\,\AA\
observations within the shaded interval by fitting an analytic
function to the observed motion. We used a Gaussian acceleration
pulse, while its actual shape is uncertain because of the double
integration in the transition from the acceleration to the
distance--time dependence. The technique to infer the kinematics is
similar to that used by \cite{Grechnev2015, Grechnev2016,
Grechnev2018}.

The initial velocity of the blob was close to zero. Its final
velocity is determined by the position of the CME core in the first
LASCO-C2 image, where it appeared; the CME frontal structure behind
the wave trace corresponds to the pre-eruption arcade enveloping the
flux-rope progenitor. The difference between the final and initial
velocities is equal to the integral over the acceleration pulse. Its
duration (and maximum) is adjusted in attempts to reproduce, on
average, the accelerating motion of the blob barely visible within
the AIA field of view.

The fit is shown in Figures \ref{F-blob}a\,--\,\ref{F-blob}e by the
black arcs. The blob underwent a maximum acceleration around
10:59:40 and reached a final speed of $1900 \pm 150$\,km\,s$^{-1}$.
The uncertainty in the duration and maximum of the Gaussian
acceleration pulse is within a factor of two.

Figure~\ref{F-blob} shows that when the acceleration ceased, the
blob lagged behind the CME leading edge measured in the CME catalog
by a factor of 1.45 at the first CME appearance in the LASCO-C2
field of view. The leading edge of the blob at that time according
to the kinematics presented in Figures
\ref{F-blob}f\,--\,\ref{F-blob}h is denoted in
Figure~\ref{F-overview}c with a dotted arc, which corresponds to the
CME core. We stated the association of an erupting flux rope with
the CME core previously that does not contradict the traditional
idea relating the flux rope to the cavity; the flux-rope forms in
the cavity from the structures of the core in the course of a
time-extended process \citep{KuzmenkoGrechnev2017}.

\subsection{Temporal Profiles of the Bursts}
  \label{S-time_profiles}

\begin{figure} 
  \centerline{\includegraphics[width=0.8\textwidth]
   {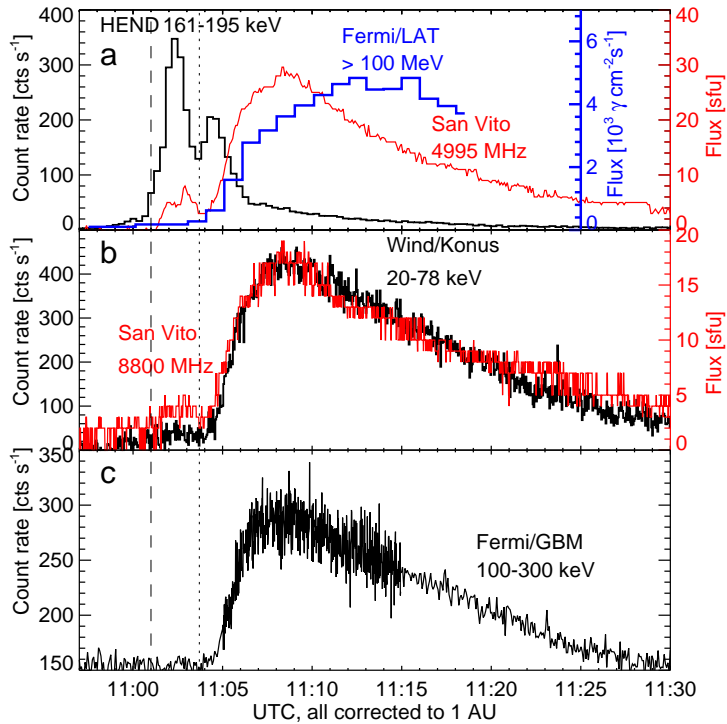}
  }
\caption{Electromagnetic emissions observed from different vantage
points. a)~Unocculted-flare HXR burst recorded from the Martian
direction by HEND (black), microwave burst at 5\,GHz recorded in San
Vito (red), and $> 100$\,MeV $\gamma$-ray burst recorded by
\textit{Fermi}/LAT (thick-blue). b)~Comparison of the HXR
(\textit{Wind}/Konus, black) and microwave (San Vito 8.8\,GHz, red)
bursts observed from the Earth's direction (similar to Figure~5 in
\citealp{Ackermann2017}). c)~Higher-energy HXR burst recorded by
\textit{Fermi}/GBM (similar to Figure~3 in
\citealp{Plotnikov2017}).}
  \label{F-timeprof}
  \end{figure}

Figure~\ref{F-timeprof} presents the bursts observed in microwaves,
HXR, and $> 100$\,MeV $\gamma$-rays from different vantage points.
Figure~\ref{F-timeprof}a shows the unocculted HXR burst recorded by
HEND in a range of 50\,--\,800\,keV with a temporal sampling of
20\,s. The HXR burst comprised two overlapping impulsive peaks, each
of about 1.5 minutes, followed by a long-lasting weaker gradual
decay. The first peak occurred around 11:02:20 and the second around
11:04:30.

The microwave burst observed by RSTN from Earth exhibits the first
minor peak corresponding to the first unocculted HXR peak around
11:02:20. The main microwave burst started nearly simultaneously
with the second unocculted HXR peak around 11:04:30 but looks
strongly ``stretched'' and lasted about half an hour. This behavior
suggests confinement of emitting electrons in a magnetic trap after
an initial impulsive injection during the second unocculted HXR
peak. The behavior of the high-energy $\gamma$-ray burst appears to
be similar to the microwave burst; it started nearly simultaneously
with the second HXR peak, being ``stretched'' still more strongly.

The microwave burst and a lower-energy HXR burst observed from the
Earth's direction by \textit{Wind}/Konus (Figure~\ref{F-timeprof}b)
were almost identical in shape, with the first minor peak and main
long-duration burst. The main HXR burst observed by
\textit{Fermi}/GBM at higher energies was similar, while the first
minor peak was indistinct.

The photon spectrum index [$\gamma$] estimated from the HEND data
was 3.27 for the first peak and 3.13 for the second peak and then
gradually hardened down to $\approx 2.2$ at 11:15, resembling the
``soft--hard--harder'' spectral behavior \citep{Kiplinger1995}. The
photon-index error caused by the dead-time correction uncertainty
does not exceed 0.3. Thus, the spectrum indices of the two
unocculted HXR peaks were almost identical. On the other hand, the
main long-duration burst visible from the Earth's direction was
obviously harder in HXR than the main burst. According to
\cite{Ackermann2017}, the emission spectrum integrated between 11:02
and 11:20 corresponded to a single power-law from 30\,keV to about
10\,MeV with an index of 2.06. This value is close to the index
estimated from HEND data for a later stage of the event.

\begin{figure} 
  \centerline{\includegraphics[width=0.8\textwidth]
   {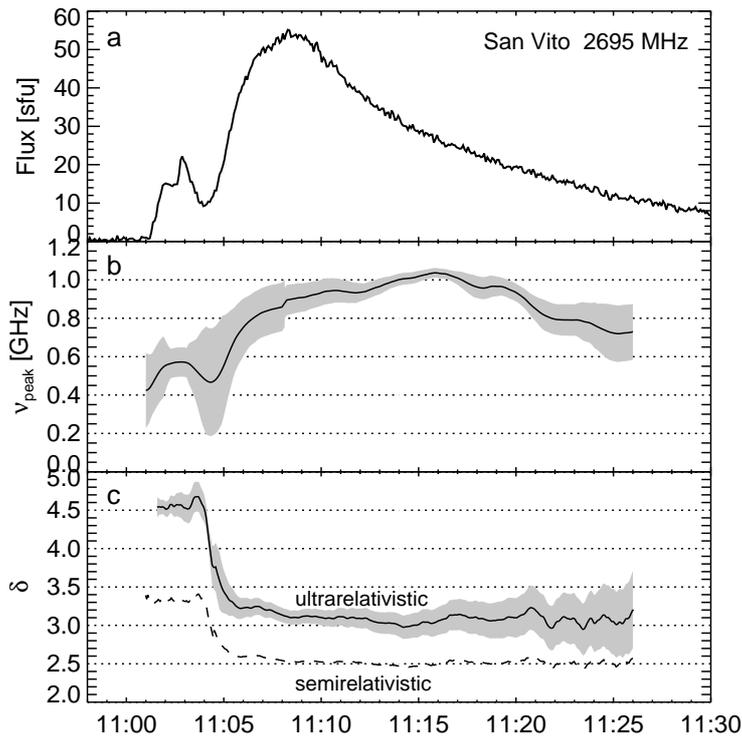}
  }
\caption{Evolution of the microwave emission during the event.
a)~Total-flux temporal profile recorded in San Vito at 2.7\,GHz.
b)~Variations of the microwave peak frequency. c)~Power-law index
of microwave-emitting electrons estimated from the slope of the GS
spectrum (ultrarelativistic limit solid, semirelativistic case
dashed). The error ranges are shown with gray shadings.}
  \label{F-fpeak_delta}

  \end{figure}

Figure~\ref{F-fpeak_delta} presents microwave spectral
characteristics in comparison with the temporal profile at
2695\,MHz. The variations in the peak frequency
[$\nu_\mathrm{peak}$] of the GS emission shown in
Figure~\ref{F-fpeak_delta}b were computed by fitting in the log--log
scale of a parabola to an instantaneous set of samples recorded at
different frequencies in San Vito (see, \textit{e.g.},
\citealp{White2003, Grechnev2013}). The shading represents the
measurement errors caused by the background-level uncertainties and
noise and does not include the calibration uncertainties, which are
not known. The peak frequency in the first minor peak was about
500\,MHz. During the main burst, $\nu_\mathrm{peak}$ increased but
did not show large variations, being within 700\,--\,1000\,MHz.

Figure~\ref{F-fpeak_delta}c presents the power-law index of the
electron energy density (electron number) spectrum estimated from
the microwave index [$\alpha$] at optically thin frequencies
considerably exceeding $\nu_\mathrm{peak}$. Usually such estimates
invoke the semirelativistic approximation by \cite{DulkMarsh1982}
(gyrosynchrotron emission), $\delta = 1.36-1.1\alpha$, where
$\alpha$ is signed and $\delta$ is always positive. The dashed line
in Figure~\ref{F-fpeak_delta}c represents $\delta$ estimated in this
way. On the other hand, according to \cite{Ackermann2017} and
\cite{Plotnikov2017}, a single-power-law electron spectrum exceeded
10\,MeV during the main burst and extended up to about 7\,MeV during
the first minor peak \citep{Carley2017}. Thus, the estimate for the
ultrarelativistic limit (synchrotron emission) might be more
applicable. In this case, $\delta = 1-2\alpha$ \citep{Dulk1985}.
Figure~\ref{F-fpeak_delta}c shows this estimate by the solid line
with uncertainties represented by the gray shading. While the
synchrotron emission matches the situation better, we use the
``gyrosynchrotron'' term following the tradition.

For the thick-target emission in the non-relativistic limit
corresponding to the HEND observations, the electron number index
$\delta = \gamma + 1.5 \approx 4.7$ ($+0.5$ relative to the
electron-flux spectrum; see \citealp{Silva2000, White2011}). This
value is close to the power-law index of microwave-emitting
electrons in the first minor peak (the GS emission is mainly
produced by the high-energy part of the electron spectrum). However,
much harder electron spectrum is suggested by the main burst.
Progressive hardening of the electron spectrum down to $\delta -
3/2$ is possible in a magnetic trap, where particles are injected
continuously \citep{MelroseBrown1976, MelnikovMagun1998,
MetcalfAlexander1999}. However, if we are really dealing here with
trapping, then the main injection was impulsive, while the
variations in $\delta$ inferred from the microwave spectrum after
11:05 seem to be too small to account for the difference between the
electron spectra in the flare HXR peak and main radio burst.

This source of the main burst was apparently different from the
source of the first minor peak. With an almost constant electron
spectrum index and nearly constant peak frequency during the main
burst, its source must be static. The gradual changes in the peak
frequency could be due to a varying number of emitting electrons and
minor variations in the electron spectrum. This behavior does not
support a possible association of the emitting source with either
the CME, whose flux-rope moved away from the solar surface up to
$\approx 3\,\mathrm{R}_\odot$ at 11:24, when the burst still
continued, or a CME-related shock wave.

To summarize, the flare was comprised of two main acceleration
episodes manifested in the HXR peaks observed by HEND. The first
episode accounts for the first minor peak of the emissions observed
from the Earth's direction. The second flare episode coincides with
the onset of the main burst in all emissions observed from
near-Earth vantage points. This burst was much longer than the flare
HXR peak that suggests a possible confinement of emitting particles
in a magnetic trap injected there during the second flare episode.
The $> 100$\,MeV $\gamma$-ray burst shows a similar behavior to the
HXR and GS burst, suggesting a common location of emitting electrons
and protons. \cite{Plotnikov2017} also concluded that accelerated
electrons and protons had a common source. The spectrum of these
particles was considerably harder than the second flare peak. If the
second flare-acceleration episode supplied particles responsible for
the main burst, then their additional acceleration was required.

\subsection{Radio Sources}
\label{S-radio_source}

\cite{Ackermann2017} showed that the radio burst was dominated by
the gyrosynchrotron emission even in the metric range, at
frequencies $>$\,200\,--\,300\,MHz. This made it possible to use
observations from the \textit{Nan{\c c}ay Radioheliograph} (NRH) in
the analysis of the GS source presented by \cite{Carley2017}. The
authors revealed an off-limb GS source with a large extent centered
above the flare position. The source appeared by 11:01 and remained
centered at this position until about 11:05, expanding along the
limb to occupy the position angles approximately from $50^\circ$ to
$87^\circ$ at 11:02. This time interval corresponds to the first
minor peak. After 11:05, the authors found a motion of the centroid
position of the source southward with a speed of $\approx
1500$\,km\,s$^{-1}$, while its height was almost unchanged. The
authors related this GS source to the CME.

The hot blob in Figure~\ref{F-blob} apparently corresponded to the
CME's flux rope, being the most probable candidate for a
CME-related GS source. However, the blob rose radially and did not
exceed laterally a narrower range of position angles from
$68^\circ$ to $78^\circ$ by 11:02. This behavior is incompatible
with that of the radio source reported by \cite{Carley2017}.

To understand the situation, we produced the images from 10-second
integrated NRH data using the \textsf{SolarSoft} NRH package
(\url{secchirh.obspm.fr/nrhpackage.php}) at 327\,MHz and 432\,MHz
during 11:00\,--\,11:10 with an integration time and steps of
30\,seconds. The images at both frequencies show that one nearly
static source (GS1) appeared at about 11:01 and faded to about
11:05, when another static source (GS2) appeared. This source was
located approximately above the Equator and had a lesser extent
along the limb. GS2 was detectable until at least 11:10.

We did not consider a lateral expansion of GS1 that
\cite{Carley2017} detected in their higher-resolution images. This
expanding component resembles in behavior the EUV wave propagation
(see the Electronic Supplementary Material
\url{AIA211_EUV_waves.mpg}) and might be due to a possible Type~II
precursor continuum or another emission not related to relativistic
electrons. Thus, the fast southward motion of the centroid position
found by \cite{Carley2017} was most likely caused by a change in the
brightness distribution among the two nearly static sources.

\begin{figure} 
  \centerline{\includegraphics[width=\textwidth]
   {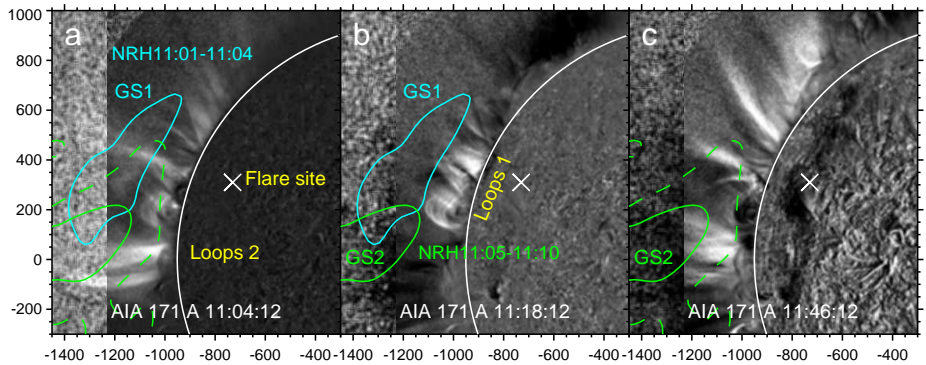}
  }
\caption{Coronal loops in AIA 171\,\AA\ and SWAP 174\,\AA\ images.
Gyrosynchrotron sources observed by NRH at 432\,MHz are shown by the
color contours. The blue contours represent the first source (GS1)
at a level of 0.6 of its maximum brightness. The green contours
represent the second source (GS2) at levels of 0.6 (solid) and 0.22
(dashed; panels a and c) of its maximum brightness. The slanted
cross denotes the projected position of the flare site.}
  \label{F-nrh_sources}
  \end{figure}

Neither GS1 nor GS2 exhibited any significant displacement, while
their possible minor motions are beyond our scope.
Figure~\ref{F-nrh_sources} presents the contours of the NRH images
averaged over the first minor peak duration for GS1 and over an
interval of 11:05\,--\,11:10 for GS2 corresponding to a considerable
part of the main burst, including its maximum. Each of the two
sources considerably exceeded the NRH beam size; thus, the NRH
images represent their realistic dimensions.

To reveal coronal structures, with which each of the two sources was
associated, we invoke the coronal-dimming phenomenon. The CME
lift-off rapidly stretches closed structures, leaving density
depletion behind it. This process shows up as dimming, whose
development is visible in the left panel of the
\url{AIA211_EUV_waves.mpg} Electronic Supplementary Material. This
panel in the movie presents the ratios of each current image to a
fixed image observed before the event. A large dimmed area expanded
in the movie. The dimming depth increased by about 11:10, and then
the coronal-plasma density started recovering. Dividing AIA images
by the deepest-dimming image at 11:10 emphasizes the coronal
structures that initially faded and then reappeared.

The AIA~171\,\AA\ image ratios in Figure~\ref{F-nrh_sources} are
shown for three times corresponding to the deepening of the dimming
in Figure~\ref{F-nrh_sources}a and its recovery in Figures
\ref{F-nrh_sources}b and \ref{F-nrh_sources}c. The field of view is
extended to the left using 174\,\AA\ images produced by the
\textit{Sun Watcher using Active Pixel System detector and Image
Processing} (SWAP: \citealp{Berghmans2006}) onboard the
\textit{Proba~2} micro-satellite.

Source GS1 was associated with an uppermost part of a far-side
arcade denoted ``Loops~1'', whose top is visible above the limb. The
near-the-limb portion of GS1 is most likely invisible because of
refraction. Source GS2 was associated with different long loops
denoted ``Loops~2'', which are deflected by the rising CME in
Figure~\ref{F-nrh_sources}a and relaxed to an equilibrium state in a
much later Figure~\ref{F-nrh_sources}c. We did not analyze a
possible small deviation in the position of GS2 that would
correspond to the minor motions of the loops. No on-disk
manifestations are visible in the NRH images, except for a
lower-frequency ($\lsim 300$\,MHz) static noise-storm source in the
southern part of the Sun that was irrelevant to the eruptive event
in question \citep{Carley2017}.

Thus, a static source GS1 was responsible for the first minor peak
around 11:02:20. \cite{Carley2017} found that this peak was caused
by the GS emission from electrons with a power-law index $\delta =
3.2$ in an energy range from 9\,keV to 6.6\,MeV in magnetic field
of $4.4$\,G and an ambient plasma density of $n_{0} = 1.3 \times
10^{8}$\,cm$^{-3}$. This power-law electron spectrum index is
close to the semirelativistic approximation in our
Figure~\ref{F-fpeak_delta}c, while the peak frequency in
Figure~\ref{F-fpeak_delta}b estimated from the San Vito data is
somewhat lower than \cite{Carley2017} found from the Sagamore Hill
fluxes reduced because of operational issues. The peak frequency
here is strongly affected by the Razin suppression, which is
determined by the ambient plasma density. It was depleted at this
time because of the developing dimming; thus, the parameters
estimated by \cite{Carley2017} are most likely correct with a
reduced ambient plasma density.

Another off-limb static source GS2 was responsible for the main
long-duration burst. To estimate its parameters, we used the GX
simulator of the GS emission \citep{KuznetsovNitaFleishman2011}. The
best fit of the actual radio spectrum near the maximum at 11:08 is
reached with a magnetic-field strength of about 1\,G, electrons with
an index of $\delta \approx 2.8$, which lies between the two
approximations in Figure~\ref{F-fpeak_delta}c, a low-energy cutoff
on the order of 100\,keV and an ambient density of a few
$10^{8}$\,cm$^{-3}$. The simulations indicate that both bases of the
emitting loops were most likely occulted; with an on-disk position
for at least one of them, the fluxes around the peak frequency
become flatter than the observations show.

There were no on-disk signatures of the GS emission. If it had been
produced by returning electrons accelerated by a shock wave, which
expanded away from the Sun (the scenario advocated by
\citealp{Plotnikov2017} and \citealp{Jin2018}), then the source
should move over the solar surface, as \cite{Hudson2017} pointed
out. This situation is not observed.

In summary, the GS sources observed by NRH confirm the indications
provided by the temporal profiles. Sources GS1 and GS2 were
distinct, each of them was nearly static and located off-limb, and
none of them was associated with the structures of the rising CME.
Source GS1 emitted by closed loops was related to the first HXR
flare peak and did not show any significant trapping. The long-lived
source GS2 appeared in a different, higher, closed structure during
the second HXR flare peak, which probably initiated the main burst
visible from the Earth's direction. The region of the GS2 radio
source is also a most probable candidate for the long-duration HXR
and $> 100$\,MeV $\gamma$-ray emissions. In contrast to the first
peak, the main burst is suggestive of a prolonged confinement of
emitting particles in a magnetic trap.

\subsection{Coronal Configuration}
 \label{S-coronal_configuration}

To analyze the coronal configuration, we compare the coronal loops
observed in the EUV with magnetic-field lines extrapolated from
photospheric magnetograms produced by the \textit{Helioseismic and
Magnetic Imager} (HMI: \citealp{Scherrer2012}) on SDO. We used the
Potential Field Source Surface (PFSS) model from the SolarSoft
package provided by the Lockheed Martin Solar and Astrophysics
Laboratory (LMSAL: \url{www.lmsal.com/~derosa/pfsspack/}). Because
the flare-hosting active region (AR) 12158 was located behind the
east limb, extrapolation is only possible from a synoptic
magnetogram. We used a magnetogram for Carrington Rotation 2155, in
which AR\,12158 was mapped about ten days after the event, being not
yet present in the previous-rotation magnetogram. Analysis of open
magnetic fields over a large part of the solar surface has led to
the results very similar to those presented by \cite{Plotnikov2017}.

\begin{figure} 
  \centerline{\includegraphics[width=\textwidth]
   {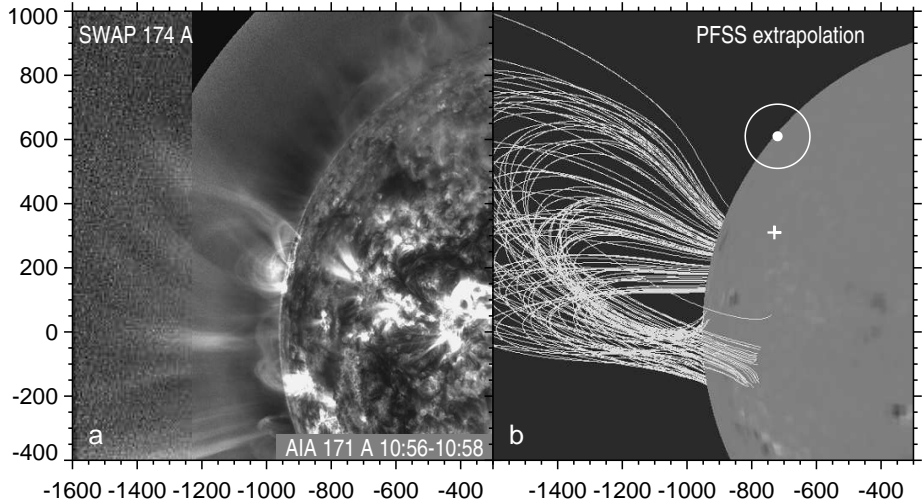}
  }
\caption{Coronal loops in a combined pre-event EUV image composed
from AIA 171\,\AA\ and SWAP 174\,\AA\ images (a) and closed
magnetic-field lines extrapolated from the flare site using the
PFSS model (b). The cross denotes the projected position of the
flare site. The dot denotes the \textit{Fermi}/LAT $> 100$\,MeV
emission centroid position with the 68\,\% error circle (from
\citealp{Ackermann2017}).}
  \label{F-euv_extrap}
  \end{figure}

For the comparison we produced a combined image of coronal loops
observed in EUV before the event. It is shown in
Figure~\ref{F-euv_extrap}a. The main part of the image is an average
over two AIA 171\,\AA\ images divided by the azimuthally averaged
radial brightness distribution (the technique is described in
\citealp{Kochanov2013}). The field of view is extended by an
enhanced-contrast average over 11 SWAP 174\,\AA\ images observed
from 10:31:21 to 10:55:12. Figure~\ref{F-euv_extrap}b presents a set
of loops extrapolated from a small region embracing the flare site.

The long loops visible in EUV near the limb close to the Equator
(Figure~\ref{F-euv_extrap}a, $y \approx 0^{\prime \prime}$) are
acceptably reproduced by the extrapolated-field lines in
Figure~\ref{F-euv_extrap}b. The lowest loops among the presented set
are comparable in length with the solar radius, and the others are
even longer. All of these loops emanate from the flare site. For
some of them, the opposite ends are rooted behind the limb, and some
others come to the visible side near the equator. Comparison with
Figure~\ref{F-nrh_sources} shows that the gyrosynchrotron source GS2
responsible for the main long-duration burst was located in these
long loops. Their Earth-facing legs visible around $y \approx
0^{\prime \prime}$ produced a brighter radio emission.

The region of the \textit{Fermi}/LAT $> 100$\,MeV centroid position
and its wide environment were entirely covered by closed field lines
over a wide range of altitudes. None of the field lines was
connected with the flare site. Neither were there any open magnetic
fields. To keep the figure decipherable, we do not show the field
lines in this region located within a very large magnetic domain
isolated from the domain in which the flare region resided.

The potential-field extrapolation used here (as well as force-free
methods) is not able to reproduce the magnetic configuration during
the CME eruption, which is the strongest violation of stationary
conditions. The violation typically involves a nearby environment of
the eruption region. In rare cases, which we call anomalous
eruptions, reconnection occurs between an erupting structure and
large-scale coronal environment (\textit{e.g.}
\citealp{Grechnev2008shock, Grechnev2011anomal, Grechnev2013neg,
Grechnev2014_I, Uralov2014}). Typical manifestations of an anomalous
eruption are dispersal of the erupted material over a considerable
part of the solar surface and microwave depressions (``negative
bursts''). Such phenomena are best visible in the 304\,\AA\ channel;
in exceptional cases they are manifested in all EUV channels
(\textit{e.g.} the SOL2011-06-07 event: \citealp{Grechnev2013neg,
vanDriel2014}). We examined all EUV channels of STEREO/EUVI and
SDO/AIA on 1 September 2014 but have not found any manifestations of
dispersed or returning erupted material. Neither there was any
microwave depression. Thus, we have not found clear support from EUV
or microwave observations to the scenario proposed by
\cite{Jin2018}. Furthermore, presumable reconnection in this
scenario between the erupting structure and the domain, where the
\textit{Fermi}/LAT $> 100$\,MeV centroid was located, had to proceed
very deep into the closed-field area to reach the connection to the
photosphere.

\section{Shock Waves}
  \label{S-shock_wave}

The presence of a shock wave in this event is indicated by a high
speed measured for the leading edge of the CME in the online CME
catalog (\url{cdaw.gsfc.nasa.gov/CME_list/}: \citealp{Yashiro2004}).
\cite{Plotnikov2017} measured some of the shock-wave characteristics
based on three-dimensional reconstructions of the wave front from
EUV and coronagraph observations. A shock wave can also be
manifested in a Type~II burst and EUV wave. As shown previously
(\textit{e.g.} \citealp{Grechnev2008shock, Grechnev2011_I,
Grechnev2016, Grechnev2017, Grechnev2018}), these signatures can be
reconciled kinematically with each other and with a halo embracing a
fast CME. Here we consider the shock-wave traces observed in EUV and
coronagraph images and possible shock signatures in a dynamic radio
spectrum.

\subsection{Shock-Wave Signatures in EUV and Coronagraph Images}
\label{S-euv_wave}

\begin{figure} 
  \centerline{\includegraphics[width=\textwidth]
   {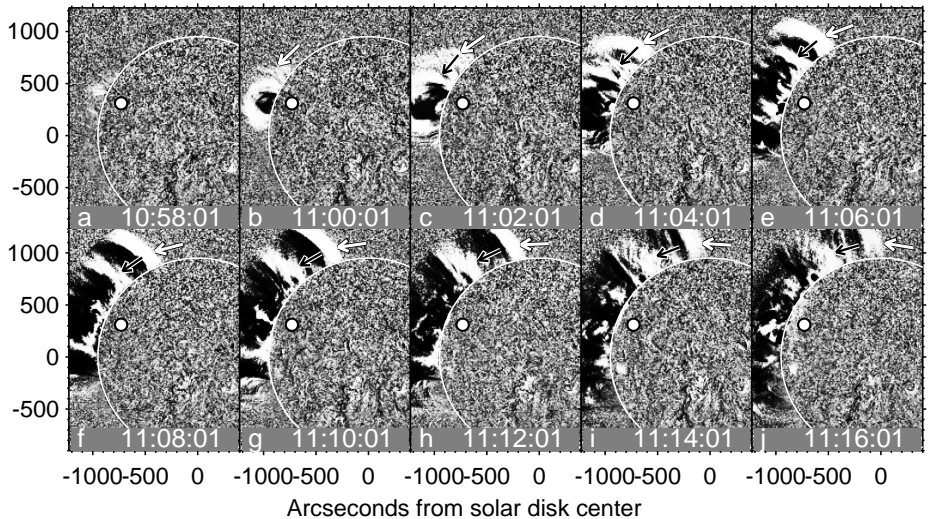}
  }
  \caption{EUV wave propagation along the limb observed in
running-difference AIA 211\,\AA\ images. The arrows point at the
first (white arrow) and second (black arrow) EUV wave fronts. The
white dot denotes the projected position of the flare site. The
axes show the distances from solar disk center in arcseconds. The
temporal interval between all consecutive images is two minutes.}
  \label{F-aia_wave}
  \end{figure}

Figure~\ref{F-aia_wave} and the Electronic Supplementary Material
\url{AIA211_EUV_waves.mpg} present EUV wave propagation observed in
AIA 211\,\AA\ images separated by two minutes.
Figure~\ref{F-aia_wave} and the right panel of the movie show
contrasted running differences. The left panel of the movie shows
the ratios of each current image with a fixed pre-event image
observed at 10:56, in which solar rotation was compensated for to
the time of the current image. Such ratio images are free from
spurious effects in running-difference images caused by subtraction.

An off-limb brightening in Figure~\ref{F-aia_wave}a facing the
far-side flare region denoted by the dot suggests that something
already happened as early as 10:58. As Figure~\ref{F-blob}
indicates, this brightening was due to expansion of high coronal
loops caused by an erupting structure, whose lift-off commenced at
that time. The EUV wave appears in Figure~\ref{F-aia_wave}b at
11:00. Its front is indicated in all images by the white arrow
parallel to the limb and by the blue bar in the movie. About two
minutes later, the second EUV wave front appears (black arrow and
red bar). Being present in non-subtracted ratio images, it cannot
be an effect of subtraction.

The northern flanks of the EUV waves manifest in a bright
compression region. The expansion at the southern flank is also
detectable, but without a clear leading brightening; it looks like
a deepening of an expanding dimming.

We measured the motion of the northern EUV waves' leading edges at a
fixed distance from the limb. The images shown in
Figure~\ref{F-aia_wave} are sampled uniformly in steps of two
minutes. The positions of the white-arrow head indicate that the
plane-of-the-sky EUV wave speed along the limb was highest initially
and then monotonically decreased. For example, the EUV wave speed in
Figure~\ref{F-aia_wave}b (11:00) was $\approx 1000$\,km\,s$^{-1}$,
and in Figure~\ref{F-aia_wave}j (11:16) it decreased to $\approx
560$\,km\,s$^{-1}$. Using a power-law fit to the measurements as
described in our articles listed in Section~\ref{S-shock_wave}, we
estimated the onset time for the first EUV wave $t_{0\,1}
=$\,10:59:04\,$\pm 15$\,seconds. It is more difficult to identify
and measure the second EUV wave front. Its probable onset time is
$t_{0\,2} =$\,11:02:00 with an uncertainty being presumably within
one minute. The measurements are presented and discussed in
Section~\ref{S-shock_measurements}; the wave speeds are shown in the
bottom panel of the movie by the corresponding colors. Both EUV
waves decelerated. Deceleration of EUV wave~2 was weaker, which is
not obvious from the plot, because the strongest-deceleration
initial part of the faster EUV wave~1 is not shown.

The EUV waves propagated over a huge area. In
Figure~\ref{F-aia_wave}j, the projected northern flank of EUV
wave~1 reached the North Pole, while the southern flank reached
the lower edge of the image shown. These moving features
apparently had a wave nature.

\begin{figure} 
  \centerline{\includegraphics[width=\textwidth]
   {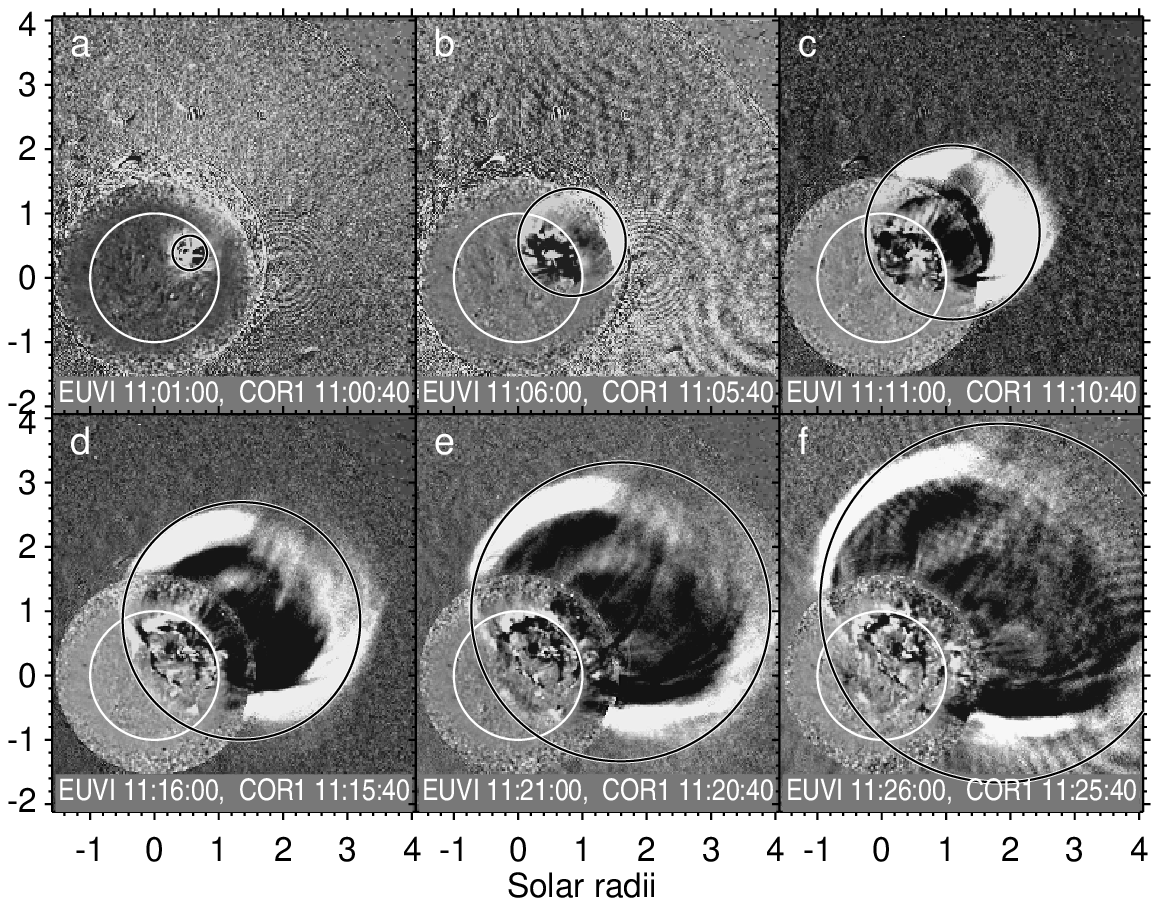}
  }
\vspace{0.1cm}
  \centerline{\includegraphics[width=\textwidth]
   {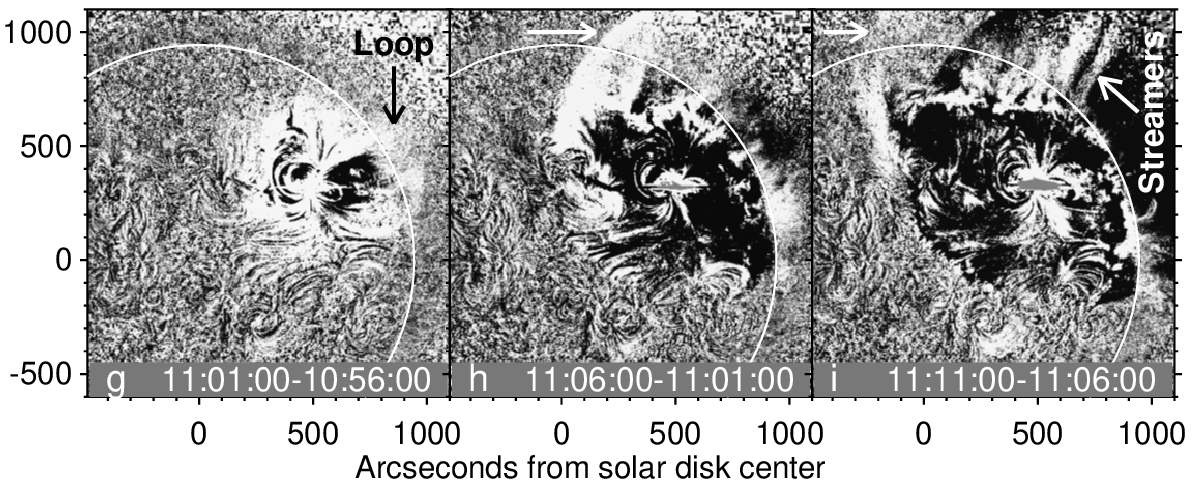}
  }
  \caption{a\,--\,f)~Wave traces observed by STEREO-B/COR1 and EUVI
195\,\AA\ running-difference images. The black-on-white circles
approximately reproduce the extent of the wave front. The white
circles denote the limb. g\,--\,i)~Wave signatures in enlarged EUVI
195\,\AA\ running-difference images shown in panels a--c. Panel g
reveals a set of long loops (``Loop''). The horizontal arrows in
panels h and i indicate the EUV wave above the limb. Panel i reveals
a set of streamer-like structures. }
  \label{F-cor1_euvi}
  \end{figure}

The expanding wave dome was also observed by STEREO-B (here we focus
on the first wave). Figures
\ref{F-cor1_euvi}a\,--\,\ref{F-cor1_euvi}f present it in combined
COR1 and EUVI 195\,\AA\ running-difference images. As
\cite{Plotnikov2017} showed, the shape of the wave front was close
to an ellipsoid (using the same method, \cite{Rouillard2016} made a
similar conclusion for a different event). Here we did not pursue to
catch the wave-dome shape; instead, the black-on-white circles in
Figures \ref{F-cor1_euvi}a\,--\,\ref{F-cor1_euvi}f approximately
reproduce its size. The correspondence between the outlining circles
and observations is almost perfect in Figures \ref{F-cor1_euvi}a and
\ref{F-cor1_euvi}b. In other panels, thick apparent flanks dominate,
being probably emphasized by deflected streamers and subtraction of
the images. Nevertheless, the circles correspond to the faint
outermost EUV wave signatures on the solar disk. These circles
corresponding to a fixed projection of an expanding ellipsoid, whose
shape does not change considerably (\textit{cf.}
\citealp{Grechnev2011_III}). Thus, the circles correctly reproduce
the kinematics of the wave-dome expansion, differing from the
highest-speed direction by a nearly constant factor. The
measurements are presented in Section~\ref{S-shock_measurements}.

Figures \ref{F-cor1_euvi}g\,--\,\ref{F-cor1_euvi}i present enlarged
EUVI 195\,\AA\ running-difference images corresponding to Figures
\ref{F-cor1_euvi}a\,--\,\ref{F-cor1_euvi}c.
Figure~\ref{F-cor1_euvi}g reveals a loop-like structure denoted
``Loop'' probably corresponding to the lower part of the loops shown
in Figure~\ref{F-euv_extrap}, in which gyrosynchrotron source GS2
was located. Figures \ref{F-cor1_euvi}h and \ref{F-cor1_euvi}i show
the EUV wave indicated by the horizontal arrow that corresponds to
Figure~\ref{F-aia_wave} viewed from a near-Earth vantage point. The
lower part of the wave front is tilted with respect to the solar
surface. Figure~\ref{F-cor1_euvi}i reveals three to five
streamer-like structures highlighted by the wave passage.

\subsection{Dynamic Radio Spectrum}
\label{S-type_ii}

\begin{figure} 
  \centerline{\includegraphics[width=\textwidth]
   {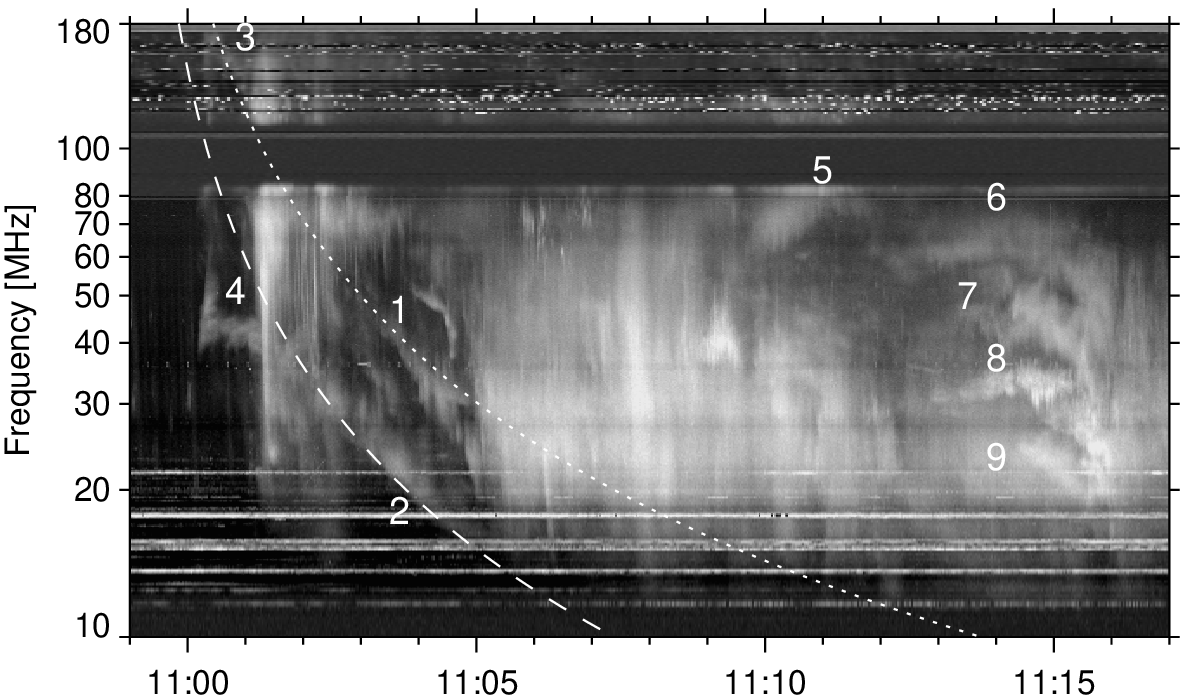}
  }
  \caption{Combined dynamic spectrum of the radio burst composed from the
NDA, Sagamore Hill, and San Vito data. The harmonically related
dotted and dashed curves correspond to an expected trajectory of a
Type~II burst produced by a shock wave with an onset time $t_{0\,1}
=$\,10:59:04 in plasma with a density falloff exponent $\mu = 2.75$.
Labels 1, 2, and 3 denote slowly drifting features that might be
possible Type~II bands. Non-drifting feature~4 with a reversely
drifting onset is also a possible Type~II-like manifestation.
Features 5\,--\,9, some with reverse drifts, might also be due to
Type~II emissions from different structures.}
  \label{F-dyn_spectrum}
  \end{figure}

A dynamic spectrum in Figure~\ref{F-dyn_spectrum} presents a radio
burst at 10\,--\,180\,MHz produced by this event. The spectrum was
composed from data of the \textit{Nan{\c c}ay Decametric Array}
(NDA: \citealp{Lecacheux2000}) at 10\,--\,80\,MHz and data of the
spectrographs at the Sagamore Hill (80\,--\,128\,MHz) and San Vito
(128\,--\,180\,MHz) RSTN stations. The dynamic spectrum presents
emissions generated at different locations. The structure of the
burst is complex and contains unusual features. Identifying Type~II
bands that carry information about a shock wave is complicated by a
series of stronger Type~IIIs (Type~VI), a gap between 85 and
110\,MHz, and interferences at higher frequencies. To search for
indications of possible Type~II bands, we plotted their expected
trajectories on top of the dynamic spectrum and, adjusting their
parameters, we tried to fit them to presumable Type~II signatures.

The method to calculate a trajectory of a Type~II burst is described
in our preceding studies (\textit{e.g.} \citealp{Grechnev2011_I,
Grechnev2017}). We use a power-law density model $n(x) =
n_0(x/h_0)^{-\mu}$ where $x$ is the distance from the eruption
center, $n_0 = 5.5 \times 10^8$\,cm$^{-3}$ is the density at a
distance $h_0 = 100$~Mm (close to the scale height), and $\mu$ is
the density falloff exponent. This model with appropriate parameters
is close to the Saito model \citep{Saito1970} in the far zone, where
the corona is quiet (Section~\ref{S-shock_measurements}), and
provides higher densities in the near zone ($< 260$\,Mm), where the
corona is strongly disturbed by the eruption. Referring to an
arbitrary point on the dynamic spectrum at time $t_1$, we choose a
frequency and calculate a corresponding distance $x_1$ from our
density model for the first or second harmonic of the plasma
frequency. The wave onset time $t_{0\,1} =$\,10:59:04 was estimated
from AIA data. Then, we calculate the Type~II trajectory as $x(t) =
x_1[(t-t_{0\,1})/(t-t_1)]^{2/(5-\mu)}$. A similar approach was used
in Section~\ref{S-euv_wave} to measure the shock-wave kinematics
from its signatures in AIA, EUVI, and COR1 images.

The dotted and dashed curves plotted in Figure~\ref{F-dyn_spectrum}
are harmonically related (2:1) and correspond to an expected
trajectory of a Type~II burst produced by the first shock wave with
an onset time $t_{0\,1}$ = 10:59:04 in plasma with a density falloff
exponent $\mu = 2.75$. The trajectories more or less correspond to
slowly drifting features 1 and 2 discernible between 11:02:40 and
11:05:00, although their structures are different. The corresponding
kinematics is presented in Section~\ref{S-shock_measurements}.

The drift rate of the faint higher-frequency feature 3 is somewhat
different from the calculated trajectories and can be reproduced by
using a lesser $\mu \approx 2.3$. The source of this emission was
probably located in a different structure.

A narrow-band ($\approx 10\,\%$) feature~4 starts from a fast
reverse drift and does not exhibit any drift afterward. Its onset
suggests an interaction of a quasi-perpendicular shock on an
extended structure with a contact point rapidly moving to its base.
This feature might be caused by a collision of a curved first shock
front with a long loop. The collision could excite wave processes
responsible for the emission at the plasma frequency in the loop.

A set of unusual features 5\,--\,9 is visible between 11:10 and
11:16. They start from reverse drifts, which later turn to the
normal direction from high to low frequencies. These features are
relatively narrow-band and have Type~III-like structure, similar to
Type~II bursts. A harmonic counterpart to feature 5 is not
detectable. The pairs (6, 8) and (7, 9) are both harmonically
related, although the structures of the bands in each pair are not
identical. Features 5\,--\,9 might also be due to Type~II emission
produced by the passage of the shock wave, but the cause of their
unusual drifts is not obvious. Their spectral evolution is different
from nondrifting Type~II-like bursts presented by \cite{Aurass2002}
and \cite{Aurass2003} and from inverse-N-like shifts of Type~II
bands \citep{Grechnev2011_I, Grechnev2014_II}.

As demonstrated in our preceding studies (\textit{e.g.}
\citealp{Grechnev2015, Grechnev2016, Grechnev2018}), the most
probable source of a narrow-band Type~II emission is a streamer. The
shock crossing the streamer deforms the plasma flow in the vicinity
of its current sheet that induces a flare-like process running along
the streamer together with the intersection point. Figures
\ref{F-cor1_euvi}c and \ref{F-cor1_euvi}i reveal a set of small
streamers visible indeed in an EUVI image at 11:11. The streamers
appeared, being blown by a shock front. An oblique shock and
compressed plasma flow behind it displace plasma in the streamer
along it. The effect is strongest near the shock normal and
decreases away from it. Thus, just after the passage of the first
shock, an inverse density distribution forms for some time in a
portion of the streamer. When the second shock hits the streamer
about three minutes later, the intersection point moves along the
streamer up. The instantaneous drift rate reflects the distorted
density distribution in the streamer caused by the passage of the
preceding shock.

The exact number of the streamers in Figure~\ref{F-cor1_euvi} is not
obvious, and neither is the exact number of the sources responsible
for features 5\,--\,9 in Figure~\ref{F-dyn_spectrum}. In addition,
the presence of a streamer is a necessary but not sufficient
condition to produce Type~II emission. With these uncertainties, the
number of the streamers roughly corresponds to the response in the
dynamic spectrum.

\subsection{Summary on Shock-Wave Measurements}
 \label{S-shock_measurements}

\begin{figure} 
  \centerline{\includegraphics[width=0.75\textwidth]
   {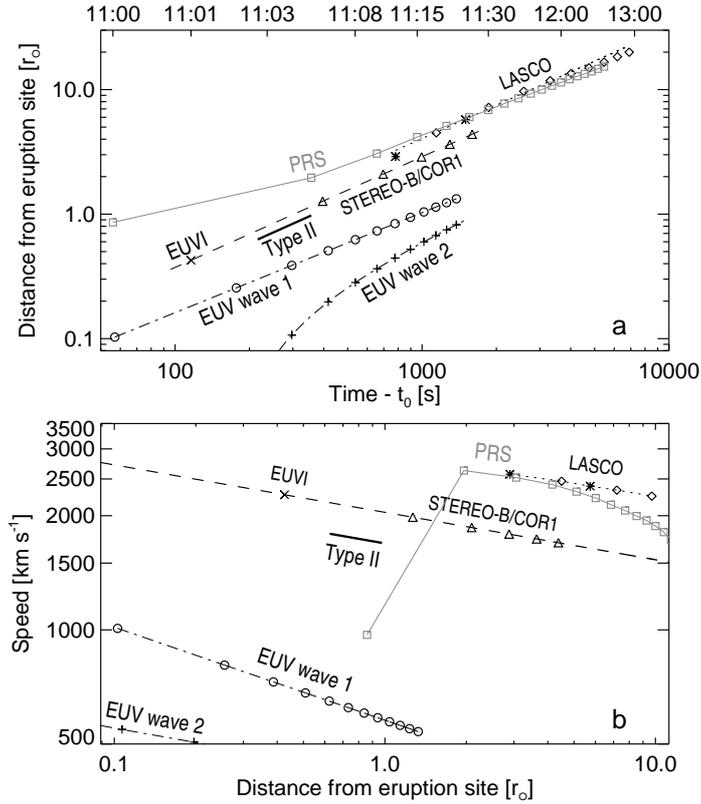}
  }
  \caption{Decelerating power-law kinematics of the shock waves
measured from different vantage points using different methods. All
distances are measured from the eruption center located at
$0.84\,\mathrm{R}_\odot$. a)~Distance--time plots. The horizontal
axis represents time after the wave onset $t_{0\,1} =$\,10:59:04 in
the logarithmic scale. The upper $X$-axis shows actual times.
b)~Speed--distance plots. The symbols represent our measurements
from the images produced by different telescopes. The measurements
from the CME catalog are shown by asterisks (C2) and diamonds (C3).
The thick bar represents the shock kinematics corresponding to
possible Type~II bands outlined in the dynamic spectrum. The gray
squares denoted ``PRS'' represent the measurements from
\cite{Plotnikov2017}. The black lines of different styles show the
analytic fit.}
  \label{F-shock-wave_plots}
  \end{figure}

The measurements in Sections \ref{S-euv_wave} and \ref{S-type_ii}
were fitted using power-law distance--time relations: $x(t) \propto
(t-t_{0\,1})^{2/(5-\mu)}$. Power-law kinematic plots of a shock wave
shown on a log--log scale with the origin of the time axis at the
wave onset time [$t_0$] are represented by straight lines. We used
in Figure~\ref{F-shock-wave_plots} the same $t_{0\,1} =$\,10:59:04
in all cases and density falloff exponents $\mu = 2.75$ for the wave
signatures in COR1 and EUVI images and dynamic spectrum;
$\mu_\mathrm{AIA} = 2.51$ for EUV wave~1, and $\mu_\mathrm{AIA} =
2.75$ for EUV wave~2, both running along the limb in AIA images
(Figure~\ref{F-aia_wave}). The plot for EUV wave~2 is conspicuously
curved, because it started three minutes after the origin of the
plot. This situation demonstrates the sensitivity of the log--log
representation to the wave onset time, which permits one to estimate
it with a high accuracy.

Figure~\ref{F-shock-wave_plots} presents the results of the
kinematical measurements and their analytic fit by different line
styles. Our measurements exactly coincide with the fitting lines,
because the outlining curves in Figures \ref{F-aia_wave},
\ref{F-cor1_euvi}, and \ref{F-dyn_spectrum} were calculated
analytically and adjusted to the observed wave signatures.

The data from the online CME catalog are shown by asterisks for the
measurements from LASCO-C2 images and by diamonds for those from C3
images. The gray squares denoted ``PRS'' represent the measurements
by \cite{Plotnikov2017}. All of these heliocentric distances are
reduced by $0.84\,\mathrm{R}_\odot$ to refer to the eruption site
(assumed to coincide with the flare position).

The measurements in the CME catalog reveal kinematics close to our
results in Figure~\ref{F-shock-wave_plots}a. The dotted line
calculated as our fit of the measurements from EUVI and COR1 images
magnified by a factor of 1.4 acceptably matches the data from the
CME catalog up to $10\,\mathrm{R}_\odot$.

All measurements in Figure~\ref{F-shock-wave_plots}a present similar
distance--time histories, except for the curved plot for EUV wave~2
because of its later wave-onset time: $t_{0\,2} > t_{0\,1}$. The
difference of $40\,\%$ between the measurements in the CME catalog
from LASCO images and our measurements from STEREO images may be
caused by the ellipticity of the shock-wave dome and still more
probably by the different techniques used. The difference within
$20\,\%$ between the measurements from COR1 images and dynamic radio
spectrum can reflect the difference in the shock-wave propagation
directions and in plasma densities in coronal structures. The
difference between the measurements from AIA and COR1 images
reflects a slower motion of the EUV wave over the solar surface and
its stronger deceleration with respect to the wave dome expanding
away from the Sun (Figures
\ref{F-cor1_euvi}b\,--\,\ref{F-cor1_euvi}f).

The slopes of the straight fitting lines are $2/(5-\mu)$
corresponding to $\mu = 2.51$ (the slope of 0.80) for EUV wave~1 and
to $\mu = 2.75$ (0.89) for all others. The power-law density model
$n(x) = n_0(x/h_0)^{-\mu}$ with $n_0 = 3.75 \times 10^8$\,cm$^{-3}$
and $\mu = 2.75$ is close to the Saito model \citep{Saito1970} at
the latitude of the flare site $14^\circ$ within $\pm\,30\,\%$ at
distances from 260\,Mm to $25\,\mathrm{R}_\odot$. Recall that we use
$n_0 = 5.5 \times 10^8$\,cm$^{-3}$ and measure the distance [$x$]
from the eruption center, while the Saito model refers to the
heliocentric distance $r = R/\mathrm{R}_\odot$, so that $x \approx
(r-1)\mathrm{R}_\odot$ in the radial direction. Overall, the
measurements made using different methods are in a reasonable
agreement with an expected propagation of a decelerating
blast-wave-like shock in a typical corona.

The expected speed--time dependencies can be obtained by
differentiation of distance--time plots, $v(t) \propto
(t-t_0)^{2/(5-\mu) - 1} = (t-t_0)^{(\mu-3)/(5-\mu)}$. The shock
wave propagating in plasma with a density falloff exponent $\mu <
3$ monotonically decelerates. We only observed decelerating shock
waves so far.

For the speed \textit{vs.} distance dependence we get from the
analytic fit $v(x) \propto x^{(\mu-3)/2}$. In this event, the
slope of the speed--distance plots is $-0.13$ ($\mu = 2.75$) for
most shock-wave signatures and $-0.23$ ($\mu = 2.54$) for EUV
wave~1. The latter value corresponds to a stronger deceleration of
the slower shock-wave trail on the solar surface in
Figure~\ref{F-cor1_euvi}, while the whole wave dome expanded
self-similarly.

The speed--distance plots are shown in
Figure~\ref{F-shock-wave_plots}b for the distances from 70~Mm to
$10\,\mathrm{R}_\odot$ from the eruption site. Our measurements from
different data agree with each other and with the measurements in
the CME catalog within this range. The results of
\cite{Plotnikov2017} within $(1.5-5)\,\mathrm{R}_\odot$ are close to
the measurements in the CME catalog, but they show a stronger
deceleration at larger distances. However, the most challenging is
their first data point with about three times lower speed than
expected, whereas our measurements show the slope persisting down to
much lower distances. The shock wave did not have any acceleration
phase. Instead, the wave started from the fast-mode speed in its
origin and monotonically decelerated. We have to conclude that the
estimate by \cite{Plotnikov2017} of the shock-wave speed at its
earliest appearance is questionable.

Our results also disagree with \cite{Jin2018}, who found the shock
speed to increase from $\approx 400$\,km\,s$^{-1}$ to $\approx
1000$\,km\,s$^{-1}$ from the tenth minute since the eruption until
the thirtieth minute. This time interval corresponds to the
measurements from STEREO-B/COR1 in Figure~\ref{F-shock-wave_plots},
where the shock speed monotonically decreases from $\approx
2000$\,km\,s$^{-1}$ to $\approx 1700$\,km\,s$^{-1}$. A possible
cause of the questionable result of \cite{Jin2018} might be the
difficulty to identify the shock front from MHD simulations that the
authors made. The incorrect behavior of the shock speed probably
affected the derived plots. We hope our results can help in
elaborating the promising method developed by the authors.

\section{Discussion}
 \label{S-discussion}

\subsection{Trapping}

One of the challenges of this event is a contrast between the
impulsive temporal profiles of the flare HXR emission observed by
HEND and long-lasting gyrosynchrotron, HXR, and $\gamma$-ray
emissions observed from the Earth's direction. Long-duration
$\gamma$-rays have been observed in the past in a few events and
extensively discussed (see, \textit{e.g.}, \citealp{Forrest1985,
Akimov1996, Ryan2000, Kurt2010, Kuznetsov2011}). One of the possible
explanations considered was long-term trapping of high-energy
protons in closed coronal loops. Microwave bursts often exhibit
manifestations of trapping of accelerated electrons (\textit{e.g.}
\citealp{MelnikovMagun1998, Silva2000, Kundu2001}). Trapping of
protons might also occur (\textit{e.g.}
\citealp{MandzhavidzeRamaty1992}). The conditions to contain trapped
relativistic protons or ions for a long time were summarized by
\cite{Ryan2000}: low density, low turbulence, and force-free field.
These requirements can be hard for flare loops but not critical for
lower-density long loops like those in Figure~\ref{F-euv_extrap} and
high-energy protons responsible for the pion-decay emission.

\subsubsection{Temporal Profiles}
 \label{S-trap_profiles}

\begin{figure} 
  \centerline{\includegraphics[width=0.8\textwidth]
   {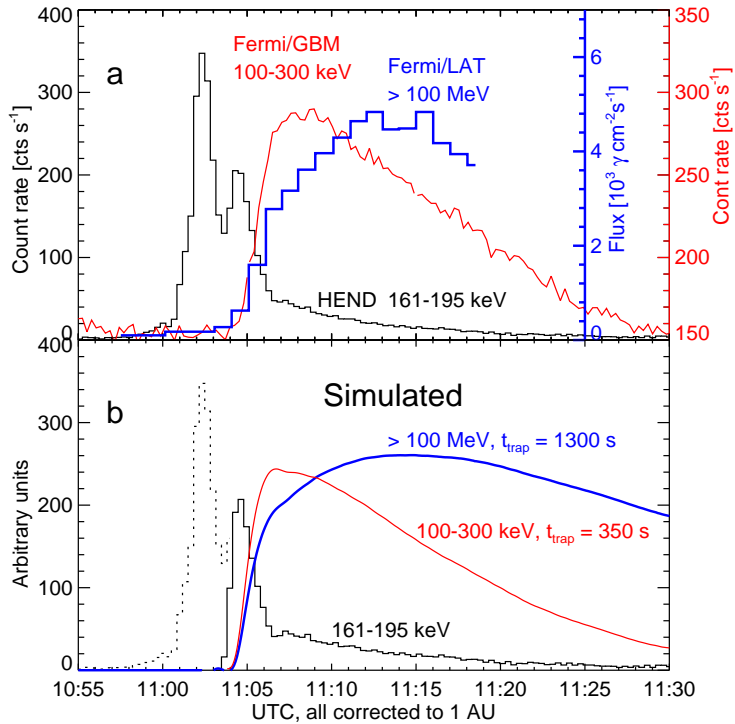}
  }
\caption{The trapping effect issue. a)~Hard emissions actually
observed (corresponding to Figure~\ref{F-timeprof}):
161\,--\,195\,keV flare hard X-rays (\textit{Mars~Odyssey}/HEND,
black), 100\,--\,300\,keV viewed from the Earth's direction
(\textit{Fermi}/GBM, red), and $> 100$\,MeV $\gamma$-ray burst
(\textit{Fermi}/LAT, thick-blue). The one-second and four-second
time-resolution \textit{Fermi}/GBM data are both summed over 16
seconds for convenience. b)~Simulated temporal profiles expected as
a result of trapping: the injection function for the second HXR peak
[$f_\mathrm{inj}$] (black-solid), actual HXR temporal profile
(black-dotted), and simulated emissions from trapped electrons (red)
and protons (thick-blue). }
  \label{F-timeprof_model}
  \end{figure}

We ask if the trapping effect can produce the observed long-duration
time profiles in response to particle populations injected
impulsively. Figure~\ref{F-timeprof_model}a reproduces the HXR and
$\gamma$-ray temporal profiles observed from different vantage
points that were presented in Figure~\ref{F-timeprof}. As noted, the
first HXR peak observed by HEND around 11:02:20 had a very close
response in gyrosynchrotron and lower-energy hard X-rays. The main
long-duration radio and HXR bursts represent the only response to
the second HEND peak around 11:04:30.

At the first step, one should separate from the actual temporal
profile the second HXR peak, which possibly was responsible for the
long-lasting emissions observed from the Earth's direction. An
appropriate shape has a function $\Psi(t, \mu, \tau) = t^\mu
\exp(-t/\tau)$ \citep{Aschwanden2004} similar to the Landau
probability distribution. To separate the second HXR peak,
decomposition of the whole HXR temporal profile is required. We
considered three peaks: a minor peak around 11:00:00, the first
peak, and the second peak. The decomposition was made by
least-squares fitting a linear combination simulating the three
peaks to the HXR$(t)$ temporal profile actually observed by HEND.
Specifically, we minimized the average value $\overline{Q}$ of the
$Q$ quantity defined as
\begin{eqnarray}
Q = \left[ \mathrm{HXR}(t) - a_0\Psi(t, \mu_0, \tau_0)- a_1\Psi(t,
\mu_1, \tau_1) - a_2\Psi(t, \mu_2, \tau_2) \right]^2.
 \nonumber
\end{eqnarray}
The dotted line in Figure~\ref{F-timeprof_model}b shows the actual
HXR temporal profile. The solid line represents the net second
peak evaluated from the decomposition.

At the second step, the emissions from trapped particles were
simulated following the approach used by \cite{Kundu2001}. The
temporal profiles [$I(t)$] were calculated as
\begin{eqnarray}
I(t) = \int_{-\infty}^{t}{\exp[-(t-t^{\prime})/\tau_\mathrm{trap}]
f_\mathrm{inj}(t^{\prime})\mathrm{d}t^{\prime}},
 \nonumber
\end{eqnarray}
where the net second peak found at the first step was used as an
injection function [$f_\mathrm{inj}$]. The trapping times
[$\tau_\mathrm{trap}$] were adjusted to make the temporal profiles
simulated more or less similar to those actually observed. The
results are shown in Figure~\ref{F-timeprof_model}b by the curves,
whose colors correspond to temporal profiles in
Figure~\ref{F-timeprof_model}a. The simulated temporal profiles
acceptably reproduce the shapes of the bursts actually observed by
\textit{Fermi} in HXR and $\gamma$-rays. We remind the reader that
the gyrosynchrotron radio burst was almost identical to the HXR
burst in shape (Figure~\ref{F-timeprof}). Thus, the simulation
demonstrates that the long-lasting HXR and radio emissions observed
from the Earth's direction can well be accounted for by trapping of
emitting electrons in closed coronal loops.

With a static position of the gyrosynchrotron source shown by NRH to
be associated with long, closed loops, its origin due to emission
from trapped electrons appears to be natural. The same is most
likely related to the HXR burst, contrary to the idea of
\cite{Plotnikov2017} about the shock-related origin of accelerated
electrons high in the corona and their return to the solar surface
along open magnetic structures. In any case, neither electrons nor
protons have access into closed loops.

\subsubsection{Low-Energy Cutoff of the Electron Spectrum}

An additional support to the role of trapping in this event is
provided by the spectral information. According to
\cite{Carley2017}, the low-energy cutoff of the electron spectrum in
the first peak during 11:01\,--\,11:04 was as low as 9\,keV. For the
main burst between 11:06 and 11:16, \cite{Plotnikov2017} estimated
the low-energy cutoff to be much higher, at 130\,keV. We obtained a
low-energy cutoff of order 100\,keV by fitting the gyrosynchrotron
spectrum near the maximum of the burst at about 11:08.

The temporal behavior of the electron energy density spectrum with
an increasing low-energy cutoff is really expected in a trap after
an initial impulsive injection, $\Delta t_\mathrm{inj} \ll
\tau_\mathrm{trap}$, of electrons with a power-law spectrum
\citep{MelroseBrown1976, MetcalfAlexander1999}. For the estimate we
use a formula from \cite{MelroseBrown1976} for the turnover energy
$E_\mathrm{T} = (3/2\nu_0 t)^{2/3}$ of the spectrum of fast trapped
electrons precipitating into the loss cone because of Coulomb
collisions, where $\nu_0 = 5 \times
10^{-9}n_0$\,[s$^{-1}$(keV)$^{3/2}$] and $n_0$ is the number density
of thermal electrons. With an impulsive injection at 11:05 and an
ambient density of $n_0 = 6.4 \times 10^{8}$\,cm$^{-3}$, we estimate
$E_\mathrm{T}$ to be 91\,keV at 11:08, 44\,keV at 11:06, and
216\,keV at 11:16. The average turnover energy $E_\mathrm{T}$
between its values at 11:06 and 11:16 is 130\,keV. These values
expected for the spectrum of trapped electrons match the low-energy
cutoffs estimated from observations. Note that our simplified
simulations and estimates in Section~\ref{S-trap_profiles} and the
present section do not consider possible changes in the plasma
density or other complications.

\subsubsection{Trapping of Protons}

The temporal profiles and spectral characteristics of the
gyrosynchrotron and HXR emissions viewed from the Earth's direction
are consistent with a long-term trapping of an electron population
impulsively injected during the second peak. As our simulation
shows, trapping of protons responsible for the pion-decay
$\gamma$-ray emission was also a likely cause of its long duration.
The $>100$\,MeV $\gamma$-ray temporal profile obtained in our
simulation acceptably matches the actual light curve. The difference
between the durations of the HXR/radio and $\gamma$-ray bursts
observed from the Earth's direction is much less than the difference
between either of them and the probable injection function.
\cite{Plotnikov2017} also concluded that accelerated electrons and
protons responsible for the long-lasting emissions in this event had
a common origin, contrary to the impression of \cite{Jin2018} about
the drastic difference between the $>100$\,MeV light curve and all
other emissions. The long-lasting $\pi^0$-decay emission observed in
a few solar events was previously considered by
\cite{MandzhavidzeRamaty1992} as evidence for trapping of particles
in solar flares.

The trapping time has extensively been discussed in the literature
(see \citealp{Aschwanden2004book} for a review and details). The
factors determining the trapping times of electrons and protons in
this event need a separate study.

\subsubsection{Appearance of Accelerated Particles in a Trap}

While the long-term trapping of electrons and protons in the long
static set of loops associated with the GS2 source appears to be
very probable, this set of loops did not exhibit direct
participation in either the flare or CME formation. This
circumstance raises a question of how high-energy particles came to
these magnetic structures disconnected from the flaring structures
and CME.

Most likely, flare-accelerated particles escape into interplanetary
space or enter magnetic traps (static or moving) due to reconnection
processes. Displacement of particles across magnetic field lines
because of drifts or collisions occurs slowly and not efficiently;
otherwise, trapping would be exceptional in solar events, contrary
to observations.

In our view, the development of an eruption and flare usually starts
within a bipolar configuration, which can be a part of a more
complex magnetic structure, \textit{e.g.} a magnetic domain of a
quadrupole configuration. In this situation, four domains share one
null point. The eruptive flux-rope carrying trapped
flare-accelerated particles moves toward this point. The passage of
the flux-rope at the null inevitably results in local reconnection
between its magnetic flux and fluxes belonging to adjacent domains.
As a result, both open and closed structures of these domains become
filled with energetic particles as well as cool plasma of the
pre-eruptive filament. Related schemes containing a single null
point were discussed by \cite{GaryMoore2004}, \cite{Masson2013},
\cite{Meshalkina2009}, \cite{Grechnev2013neg}, and
\cite{Uralov2014}. Stretching a large-scale quadrupole into the
solar wind might cause disappearance of the null point.
Nevertheless, lateral reconnection between the flux-rope and coronal
rays also occurs in this situation (\textit{e.g.}
\citealp{Bemporad2010}). The presence of a coronal null in the
parent active region 12158 is supported by an S--N--S--N
configuration discernible in SDO/HMI magnetograms observed one week
after the event.

\subsection{Position of the Gamma-Ray Source}
  \label{S-gamma-ray_source}

The $> 100$\,MeV emission centroid position calculated by
\cite{Ackermann2017} from \textit{Fermi}/LAT data with a 68\,\%
error radius of $100^{\prime \prime}$ is located at the east limb at
a latitude of about N41 (Figures \ref{F-overview}b and
\ref{F-overview}c). As noted in
Section~\ref{S-coronal_configuration}, this site and its wide
environment were totally covered by closed magnetic fields. Protons
and other charged particles did not have access into this domain
from either the flare region or larger coronal heights along open
field lines. On the other hand, the off-limb radio source GS2, which
is the most probable candidate for the source of long-duration HXR
and $\gamma$-ray emissions, was located nearly above the Equator,
far away from the \textit{Fermi}/LAT $> 100$\,MeV emission centroid
position. No other candidate for the source of the long-duration
burst was found. The difference between the expected position of the
$\gamma$-ray source and the $> 100$\,MeV emission centroid position
computed by \cite{Ackermann2017} induces thinking about its possible
causes.

The centroid position of the observed $\gamma$-ray emission can be
due to superposition of a few different sources. For example,
high-energy cosmic rays arriving from outside the solar system can
produce cascades of secondary particles and $\gamma$-rays in the
solar corona. Next, intense fluxes of energetic particles and
emissions from the flare site can interact with dense streamers,
producing secondaries, in particular $\gamma$-rays. Furthermore, the
expanding CME is a low-density but huge target for both extra-solar
cosmic rays and energetic flare emissions. \cite{KahlerRagot2008}
showed a possibility for high-energy $\gamma$-rays to be produced
even in interactions between SEPs and solar wind. All of these
presumable processes might influence the centroid position, while
their effects are mostly expected in the lowest-energy part of the
spectrum observed by \textit{Fermi}/LAT.

Proceeding from these considerations, we attempted to find a
possible dependence of the \textit{Fermi}/LAT emission centroid
position on the low-energy threshold. We analyzed the Level 1 Photon
File available at \url{fermi.gsfc.nasa.gov/ssc/data/} that included
the SOL2014-09-01 event. The file presents the time, energy, and
position (arrival direction) measured for each individual
$\gamma$-ray photon out of numerous discrete sources detected during
the observational interval recorded in the file. We calculated the
emission centroid positions from these data, eliminating the
$\gamma$-ray photons with energies below a given threshold
$E_{\min}$. We did not reproduce the sophisticated method used by
\cite{Ackermann2017} to reach the highest accuracy, making the
calculations in the same way for each $E_{\min}$.

The centroid position that we found with $E_{\min} = 100$\,MeV was
close to the result of \cite{Ackermann2017}. Then we increased
$E_{\min}$ in steps of 50\,MeV and found a monotonic displacement of
the centroid position along the limb toward the southeast. The
increase of $E_{\min}$ from 100\,MeV to 300\,MeV shifted it by
$\approx 30^{\circ}$ toward the radio source GS2. Our experiment
shows that the effects mentioned in this section can account for the
discrepancy between the expected position of the source and the
centroid position actually measured. This issue needs further study.

The energy dependence of the $\gamma$-ray centroid position is
difficult to reconcile with the scenario proposed by \cite{Jin2018}.
We also recall the similarity of simulated HXR and $\gamma$-ray
temporal profiles emitted from the trap after the same impulsive
injection with those actually observed
(Section~\ref{S-trap_profiles}). To fit within the scenario by
\cite{Jin2018}, accelerated electrons and protons of a common origin
have to be separated and enter different structures to emit at the
positions located far away from each other. Electrons have to come
to the off-limb source GS2, while protons have to precipitate at the
on-disk $\gamma$-ray centroid position. It seems difficult to
realize this separation. GS2 appears to be a more probable source of
both HXR and $\gamma$-ray emissions. According to \cite{Hudson2017},
the column density $nL$ required for the effective interaction of
high-energy protons with ambient plasma can be accumulated in their
numerous flights in a coronal trap (large $L$) and not necessarily
be due to a large $n$ in the photosphere.

\subsection{Histories and Possible Roles of Shock Waves}
  \label{S-shock_history}

\subsubsection{Excitation and Evolution of Shock Waves}

As shown in our preceding case studies of shock-wave histories in a
dozen events, the only initial shock-wave excitation scenario
observed in flare-related eruptions is the impulsive-piston
mechanism \citep{Grechnev2018}. Here, a sharply erupting flux rope
produces strong MHD disturbance, whose initial speed is determined
by the fast-mode speed $V_\mathrm{fast}$, which is high in active
regions ($V_\mathrm{fast} > 1000$\,km\,s$^{-1}$). Away from the
eruption site, the $V_\mathrm{fast}$ in the environment decreases
both upwards and laterally, reaching $\approx 200$\,km\,s$^{-1}$
above the quiet Sun. When a high-speed disturbance enters the
lower-$V_\mathrm{fast}$ environment, its profile steepens, and the
disturbance rapidly becomes a shock wave. The shock formation is
governed by the maximum acceleration of the eruption and the
$V_\mathrm{fast}$ falloff away from the eruption site, occurring
presumably in 10\,--\,100\,seconds
\citep{AfanasyevUralovGrechnev2013}. Then the shock wave propagates
quasi-freely up to considerable distances from the Sun like a
decelerating blast wave. Being highly efficient, the
impulsive-piston scenario initially precedes the bow-shock
excitation by the outer surface of a CME that is only possible when
it becomes super-Alfv{\' e}nic. The change to the bow-shock regime
occurs later, if the trailing CME is fast \citep{Grechnev2015,
Grechnev2017}.

The onset time of a shock wave excited in this way falls in the
acceleration phase of the responsible eruption, \textit{i.e.}
during the rise phase of an HXR (microwave) burst. In a number of
events, the acceleration of an eruption and shock onset time
advanced the bursts by about two minutes (\textit{e.g.}
\citealp{Grechnev2011_I, Grechnev2013, Grechnev2015, Grechnev2016,
Grechnev2018}). In several events, two shock waves excited within
a few minutes by different eruptions were observed to follow each
other. As shown in the articles listed, flare-generated shock
waves are unlikely.

These conclusions shed light on the event in question. The presence
of two EUV waves with onset times at 10:59:04 and 11:02:00 indicates
excitation of two shock waves one after another by two presumable
eruptions responsible for the HXR peaks observed by HEND around
11:02:20 and 11:04:30. Note that two bow shocks cannot be driven by
a single CME. Most likely, two shock waves following each other
eventually merge into a single, stronger shock propagating nearly
radially \citep{Grechnev2011_I, Grechnev2013, Grechnev2017}. We do
not have sufficient data about this process in our event and
consider here a single shock wave relating it to the first one.

In the power-law description of a shock wave, the plasma density and
wave speed become infinite in the origin ($t = 0, x = 0$). This
singularity is not important, because the shock forms at a certain
distance from the origin, while the initial wave speed is determined
by the fast-mode speed. From Figure~\ref{F-shock-wave_plots}b, the
initial shock-wave speed in the radial direction was roughly about
3500\,km\,s$^{-1}$, which is a normal value for $V_\mathrm{fast}$ in
an active region. The kinematical histories of the shock waves at
least up to $10\,\mathrm{R}_\odot$ exhibit an overall quantitative
agreement with the expected evolution outlined in the preceding
paragraphs. Here we did not follow the shock-wave evolution in
coronagraph images; the close correspondence of the calculated wave
fronts to their signatures in the images and agreement with the
measurements in the CME catalog was shown for several events
previously \citep{Grechnev2011_I, Grechnev2011_III, Grechnev2014_II,
Grechnev2015, Grechnev2016, Grechnev2017, Grechnev2018}.

There is no reason to presume the 1 September 2014 event to be
exceptional. The shock-wave excitation and subsequent evolution
appear to correspond to the impulsive-piston scenario outlined
above. This shock-wave history turns out to be more complex than
the bow-shock excitation by a super-Alfv{\' e}nic piston alone,
being, in fact, a combination of known scenarios (see,
\textit{e.g.}, \citealp{VrsnakCliver2008}). Missing this
circumstance can result in incorrect estimates and inadequate
conclusions.

In this respect, a question remains about the first estimate by
\cite{Plotnikov2017} of the shock-wave speed, which is an outlier in
Figure~\ref{F-shock-wave_plots}. It follows from the description of
the method in \cite{Rouillard2016} that the speed is calculated from
the spatial separation of successive shock-front ellipsoids obtained
in the three-dimensional reconstruction that appears to be
justified. However, this method does not provide an estimate for the
first and last speeds. Probably, this is the key to the problem,
which seems to be systematic; the initial speeds also seem to be
strongly underestimated for the three different events addressed in
these articles. Another possible source of an additional error can
be an apparent temporal difference between the first SDO/AIA images
and a STEREO/EUVI or COR1 images presented in the articles;
combination of the highest shock speed with the smallest size of its
front can result in a large error for the initial point.

On the other hand, we emphasize the importance of a particular
result of \cite{Rouillard2016} and \cite{Plotnikov2017} about the
ellipsoidal shape of the shock front that, in fact, confirms the
scenario outlined above. A similar shape of the shock-wave front
was predicted by \cite{Grechnev2011_I}, contrary to the bow shape
with a Mach cone considered by \cite{OntiverosVourlidas2009}. The
reason is a three-dimensional expansion of the CME body
(\textit{cf.} \citealp{VrsnakCliver2008}). Our expectations were
later confirmed in studies by \cite{Kwon2014} and \cite{Kwon2015}.
Elaboration of the shock-front reconstruction method by
\cite{Rouillard2016} promises further progress in understanding
coronal shock waves.

\subsubsection{Possible Particle Reacceleration by an Oblique Shock Wave}

As \cite{Hudson2017} noted, our event resembles the SOL1969-03-30
event addressed by \cite{FrostDennis1971}, who considered the HXR
emission observed in that event as evidence for two-stage electron
acceleration (initially assumed by \citealp{Wild1963}). The
first-stage acceleration was associated with flare processes, while
the shock front could be responsible for the second-stage
acceleration. The two-stage acceleration can also apply to protons.
Observations of the SOL2014-09-01 event might shed light on this
issue.

An additional acceleration of high-energy protons in a static
magnetic trap could be caused by a fast magnetosonic shock wave,
whose front positions are shown in Figure~\ref{F-aia_wave}. The
shock front propagating with the phase velocity $V_\mathrm{sh}$
intersects a part of the magnetic trap at an angle [$\psi$] to the
magnetic field [$\textbf{\textit{B}}$]. An important characteristic
here is the velocity [$u$] of the intersection point between a
magnetic field line and the shock-front surface: $u = V_\mathrm{sh}
\tan \psi$. We briefly discuss the case of an oblique shock wave
with $u < c$.

High-energy particles with gyroradii considerably exceeding the
shock-front thickness change their energy by virtue of the following
effects. The first effect results from the first adiabatic-invariant
$p_{\perp}^2/B$ conservation, where $p_{\perp}$ is a component of
the particle momentum perpendicular to the magnetic field $p  = mV /
(1 - V^2/c^2)^{1/2}$, $m$ is the rest mass of the particle, and $V$
is its velocity. Particles that are initially upstream of the shock,
with pitch angles in the loss cone, pass into the downstream region
of a stronger magnetic field and increase their transverse kinetic
energy $K_\perp$. On the other hand, their longitudinal kinetic
energy $K_\parallel$ also changes because of the second effect of
bouncing particles against the moving magnetic mirror of the shock
front. The change $\Delta E = \Delta K$ in the total energy $E =
mc^2 + K_\perp + K_\parallel = mc^2 + K$ depends on the initial
pitch angle, velocity $u$, and $\psi$, and it can be either positive
or negative.

The growth of the total energy due to the increasing $K_\parallel$
is most conspicuous for particles reflected upstream after their
interaction with the shock. In this case, $\Delta K =
2up'/(1-u^2/c^2)^{1/2}$ \citep{Webb1983}, where $p'$ is a
gyrophase-averaged value of the particle's momentum projection on
the shock-front surface in the frame moving along the front with the
velocity [$u$]. In the non-relativistic limit ($u^2/c^2 \ll 1$),
$\Delta K = 2u(2mK')^{1/2}/(1-u^2/c^2)^{1/2}$, where $K' =
(p')^2/2m$. The particles gain energy owing to the reflection from a
moving magnetic mirror, \textit{i.e.} the shock front for upstream
particles with sufficiently large pitch angles. We are only
interested in a qualitative analysis of the particle's behavior in a
magnetic trap, through which a shock wave passes. Therefore, we
replace the last relation with a rough proportion $\Delta
K_\parallel \propto K_\parallel^{1/2}$. The higher the energy of a
particle, the larger an increase in its energy and velocity per each
reflection. The higher the particle velocity, the more reflections
it has in bouncing between the moving shock front and the base of
the magnetic trap. Thus, particle acceleration is accompanied by
flattening of the initial energy spectrum.

After the shock-front passage, the magnetic-loop trap compresses and
displaces following the wave. The magnetic field strength $B$
increases. Then the trap returns to its initial condition, and $B$
decreases to the initial value. With a decreasing $B$, the
transversal energy $K_\perp$ of each particle decreases
approximately as much as it increased in the interaction with the
shock front because of the first adiabatic-invariant conservation.
However, the longitudinal energy $K_\parallel$ accumulated in the
reflections from the shock front is retained, as well as the
energy-spectrum flattening. This conclusion seems to correspond to
the observations.

The completeness of this scheme for the proton acceleration in a
trap is open to question. The acceleration mechanism based on
reflections from the magnetic mirror in the shock front leads to an
increase in the longitudinal energies of particles. This suggests a
decrease in their pitch angles and possible precipitation into the
loss cone of the magnetic trap. Precipitation of a fraction of
energetic particles into the bases of the trap is expected to occur
in the course of the oblique-shock propagation through a trap. This
phenomenon might be manifested in the long-duration tail of the HXR
emission observed by HEND in Figure~\ref{F-timeprof}a. Precipitation
may be reduced because of the electric field originating due to the
charge separation in the front of a collisionless shock wave. Such
electric field increases the transverse energy of reflected protons
to prolong their confinement in the trap.

\section{Summary and Conclusions}
 \label{S-conclusion}

A combined analysis of observations of the far-side SOL2014-09-01
event from different vantage points has revealed the following
circumstances.

\begin{enumerate}[i]

 \item
The lift-off of a hot (about 10\,MK) blob has been detected, which
probably was an erupting flux rope. The blob rose radially and
became the CME core.

 \item
Unocculted flare emission consisted of two HXR peaks with similar
spectra separated by 2.5 minutes.

 \item
Each of the two flare peaks was preceded by the appearance of a
shock wave by two to three minutes.

 \item
The first HXR peak was manifested in a static off-limb
gyrosynchrotron radio source of a corresponding duration and
spectrum.

 \item
The second HXR peak gave rise to a different static off-limb
gyrosynchrotron radio source of a considerably longer duration and
harder spectrum. This radio source was located in a system of long
loops.

 \item
The long-duration gyrosynchrotron burst from the second source was
almost identical in shape with the HXR burst observed from the
Earth's direction and rather similar to the $> 100$\,MeV
$\gamma$-ray burst. All of these emissions could be produced by
populations of electrons and protons injected into the long loops
during the second flare burst and trapped there for a long time.

 \item
The harder spectrum of the long-duration burst relative to the
injection could be due to reacceleration of the particles trapped
in closed loops by the second shock wave.

 \item
The observations indicate that the sources of the gyrosynchrotron,
HXR, and $\gamma$-ray emissions had a common location. It was
considerably displaced with respect to the $> 100$\,MeV $\gamma$-ray
emission centroid position. A probable key to the discrepancy is a
contribution of $\gamma$-rays coming from high coronal structures
and possibly the CME. The role of non-solar high-energy cosmic rays
is not excluded.

\end{enumerate}

These findings can be reconciled within the following scenario. Two
sharp eruptions probably occurred in Active Region 12158 with an
interval of about 2.5 minutes. Each eruption impulsively excited a
blast-wave-like shock, on the one hand, and resulted in strong
particle acceleration in the flare site, on the other hand.
Manifestations of the first flare peak were observed from the
Earth's direction as an impulsive brightening of the arcade top.
During the second peak, accelerated electrons and protons were
injected into long loops, where they become trapped for a long time.
The second shock wave possibly hit these loops obliquely, which
resulted in reacceleration of trapped flare-accelerated electrons
and protons. This presumable episode was not crucial; the
long-duration gyrosynchrotron, hard X-ray, and $\gamma$-ray
emissions were radiated from trapped particles, while reacceleration
hardened their spectra. A presumable scenario with a
shock-acceleration of particles high in the corona and their return
to the solar surface along open magnetic structures meets basic
difficulties and is not confirmed by observations.

The region of trapped electrons and protons was located above the
limb. Its connection to the Earth-facing solar surface near the
Equator is not excluded, but does not seem to be necessary.

While our analysis sheds additional light on this event, a number of
issues remain to be addressed. We hope that our results will
highlight possible ways for future studies.

\begin{acks}

This work is dedicated to the memory of M.A.~Livshits, who initiated
this study. We appreciate discussions with E.~Carley, N.~Vilmer, and
H.~Hudson, and useful remarks of the anonymous reviewer. We thank
the NASA/SDO and the AIA and HMI science teams; the NASA's
STEREO/SECCHI science and instrument teams; the teams of the SWAP
telescope on the ESA's PROBA~2 spacecraft, the NASA's \textit{Fermi
Gamma-Ray Space Telescope}, the \textit{Wind}/Konus team at the
Ioffe Institute, the USAF RSTN network, and LASCO on SOHO. SOHO is a
project of international cooperation between ESA and NASA. We thank
the team maintaining the CME Catalog at the CDAW Data Center by NASA
and the Catholic University of America in cooperation with the Naval
Research Laboratory.

The studies presented in Sections \ref{S-introduction},
\ref{S-overview}, and \ref{S-time_profiles} were carried out by
V.~Kiselev and I.~Grigorieva and funded by the Russian Foundation of
Basic Research under grant 17-32-50040\_mol\_nr. V.~Grechnev,
A.~Kochanov, and A.~Uralov (Sections \ref{S-shock_wave},
\ref{S-discussion}, and \ref{S-conclusion}) were funded by the
Russian Science Foundation under grant 18-12-00172.

\end{acks}

\section*{Disclosure of Potential Conflicts of Interest} The authors
claim that they have no conflicts of interest.

\bibliographystyle{spr-mp-sola}

\bibliography{2014-09-01}

\end{article}

\end{document}